\documentclass[aps,twocolumn,showpacs,floats,prb]{revtex4}
\usepackage{epsf}
\usepackage{float}
\usepackage{amsmath}
\usepackage{graphicx}

\newcommand{\mr}[1]{{{\mathrm{#1}}}}
\newcommand{\mcal}[1]{{\mathcal{#1}}}

\newcommand{\w}{\omega}
\newcommand{\s}{\sigma}

\newcommand{\Simp}{S_\mr{imp}}
\newcommand{\Cimp}{\chi_\mr{imp}}
\newcommand{\Cloc}{\chi_{\mr{loc}}}
\newcommand{\T}{\mathcal{T}}
\newcommand{\AAA}{\mathcal{A}}

\begin{document}

\title {
Universal crossovers and critical dynamics of quantum phase transitions:\\
A renormalization group study of the pseudogap Kondo problem
}
%Universal crossovers and critical dynamics of the pseudogap Kondo problem:\\
%a renormalization group study}

\author{Lars Fritz$^{1,2}$, Serge Florens$^1$, and Matthias Vojta$^2$}
\affiliation{\mbox{$^1$Institut f\"ur Theorie der Kondensierten Materie,
Universit\"at Karlsruhe, 76128 Karlsruhe, Germany}\\ \mbox
{$^2$ Institut f\"{u}r Theoretische
Physik, Universit\"{a}t zu K\"{o}ln,
    50937 K\"{o}ln, Germany}}

\date{October 16, 2006}

\begin{abstract}
The pseudogap Kondo problem, describing a magnetic impurity
embedded in an electronic environment with a power-law density of states,
displays continuous quantum phase transitions between free and screened moment
phases.
In this paper we employ renormalization group techniques to analytically calculate
universal crossover functions, associated to these transitions, for various
observables.
Quantitative agreement with the results of Numerical Renormalization
Group (NRG) simulations is obtained for temperature-dependent static and
zero-temperature dynamic quantities, at and away from criticality.
In the notoriously difficult realm of finite-temperature low-frequency dynamics,
usually inaccessible to both NRG and perturbative methods, we show that
progress can be made by a suitable renormalization procedure in the
framework of the Callan-Symanzik equations.
Our general strategy can be extended to other zero-temperature
phase transitions, both in quantum impurity models and bulk systems.
\end{abstract}

\pacs{75.20.Hr,74.70.-b}

\maketitle

%%%%%%%%%%%%%%%%%%%%%%%%%%%%%%%%%%%%%%%%%%%%%%%%%%%%%%%%%%%%%%%%%%%%%%%

\section{Introduction}

Continuous quantum phase transitions\cite{book} are a subject of intense research,
both in experimental and theoretical condensed matter physics.
Particularly interesting are the low-temperature, low-energy properties
near quantum critical points, which are crucially influenced by the
non-trivial properties of the quantum critical ground state.
A theoretical description requires the knowledge of the low-energy
long-wavelength effective theory for the critical degrees of freedom.
Critical exponents, characterizing the power-law behavior of observables upon
approaching the critical point, may then be obtained via suitable
renormalization group (RG) techniques.
A related, but considerably harder, task is the calculation of full
crossover functions describing the universal behavior of observables,
e.g., along the changes that take place from the high-temperature 
quantum critical region to the low-temperature regimes, which appear 
in the vicinity of simpler (ordered or disordered) quantum mechanical 
ground states.
These crossover functions are particularly relevant for experiments,
as they allow to analyze measurements in terms of data collapse and
universal scaling.

On theoretical side, a large body of work has concentrated on quantum critical
points above their upper critical dimension where observables can be calculated in
straightforward perturbation theory, but where universality is limited due to
ultraviolet (UV) divergencies and the absence of hyperscaling.
In contrast, interacting quantum critical points below their upper critical
dimension require a more sophisticated treatment, and relatively little is known
regarding full crossover functions.
Exceptions are models for insulating magnets, with many results summarized in
Ref.~\onlinecite{book}.

An interesting class of model systems showing quantum phase transitions,
with both interacting and non-interacting critical points,
are quantum impurity models.
The transitions here are conceptually simpler than in bulk systems, as
only a finite number of degrees of freedom becomes critical at the
transition point.\cite{vojta_review}
Powerful numerical techniques, most notably Wilson's Numerical Renormalization
Group\cite{wilson} (NRG),
have been developed to study the asymptotic low-energy regime of impurity models,
and thus allow to benchmark analytical approaches.
Besides being model systems for fascinating many-body physics,
including the Kondo effect and local non-Fermi liquid behavior,\cite{hewson}
quantum impurity models are of great relevance for nanofabricated quantum
dots as well as for correlated bulk quantum systems treated within
dynamical mean-field theory\cite{DMFT} (DMFT) and its generalizations.

The so-called pseudogap Kondo problem\cite{fradkin,ingersent} displays
particularly rich physics including various impurity quantum phase transition.
Here, a quantum spin 1/2 couples to conduction electrons with a semi-metallic
density of states (DOS) that vanishes as $\rho(\omega) \propto |\omega|^r$ ($r>0$)
around the Fermi level.
Although only special values of the pseudogap exponent $r$ can be
experimentally accessed to date
(namely $r=0$ in metals, $r=1$ in graphene layers and $d$-wave superconductors,
and $r=2$ in $p$-wave superconductors), this model in its generality displays
a challenging variety of non-trivial properties.
%whose full understanding is a challenge for the theory.
%It is important to underline that technical methods and physical ideas developed for this
%special and simplified model are certainly useful in more complicated
%contexts.
Recently, it was realized that {\it all} critical fixed points of
the pseudogap model can be accessed perturbatively after
having identified the values of $r$ that correspond to
(upper or lower) critical dimensions of the problem.\cite{fritz2,fritz1}
Subsequently, critical properties of the model have been analytically calculated using
perturbative RG techniques.
The results for various critical exponents and {\it static} observables
{\it at criticality} (such as the entropy or the anomalous Curie constant)
show impressive agreement with NRG simulations.\cite{fritz1}
This approach illuminates the need to consider the adequate degrees of freedom to describe
a given quantum phase transition (see Ref.~\onlinecite{USPG} for a simple
illustration).
%On a technical level, an elegant formulation of the field-theoretic renormalization
%group works extremely well for this problem, and can also be employed to study more
%complex models, such as the Bose-Fermi Kondo model~\cite{si1,zarand1}.

The purpose of this paper is to extend the analytical approach of Ref.~\onlinecite{fritz1}
to the calculation of full crossover functions for static {\it and} dynamic
quantities, {\it both} at and away from criticality.
Technically, the calculation of the scaling behavior of the physical quantities
will be achieved by using the powerful Callan-Symanzik equations\cite{bgz,peskin,zinn}
within the field-theoretic RG, allowing us to compute the universal crossovers functions 
of the most interesting physical observables within the renormalized theory.
We will restrict our attention to the particle-hole symmetric case
and to the range of the DOS exponent $r=0\ldots 1/2$,
but our techniques can be applied to the particle-hole asymmetric situation\cite{fritz2}
near $r=1$ as well.
Comparison with numerical data from NRG will be made,
with extremely good agreement regarding both the temperature dependence
of static observables, and the zero-temperature dynamic correlation
functions.
With respect to the finite-temperature dynamics, it is known that direct
perturbation theory displays unphysical singularities in the low-frequency
regime.\cite{sachdev}
We show, however, that the use of Callan-Symanzik equations
may allow some progress in obtaining physically sound results.
Because the regime where frequency is smaller than temperature is also notoriously hard
to access via NRG simulations,\cite{nrg_t} our results could possibly
be useful in the future for benchmarking the quality of this numerical approach.
In a more general context, we believe that the powerful methodology employed here
(being relatively unfamiliar in the condensed matter context)
will also be of interest for general studies of quantum phase transitions, e.g.,
in quantum impurity models or bulk magnets.

We note that some scaling functions to be determined below
can also be obtained within the large-$N$ expansion of multichannel Kondo models,
both for the metallic\cite{parcollet,florens} and pseudogap\cite{vojta}
density of states.
In addition, a partial study of the dynamics at the critical point using
renormalization ideas was undertaken in Ref.~\onlinecite{si2},
and the local moment approach\cite{logan} was also applied to this problem
and compared to NRG.\cite{bulla}
We will comment on these results in the body of the paper.

The bulk of the paper is organized as follows:
In Sec.~\ref{model} we briefly review the physics of the pseudogap Kondo model,
in particular the possible quantum phase transitions and the associated
observables.
Focusing first on the weak-coupling regime of the pseudogap Kondo model, which
is perturbatively controlled for small values of $r$, we develop in
Sec.~\ref{kondo} the necessary formalism to compute the full universal
crossovers functions of the observables, and compare the analytic
results to NRG data.
In Sec.~\ref{anderson} we address the regime where $r$ is not small, using a previously
developed effective theory,\cite{fritz1} valid when $r$ is close to the value 1/2.
The calculation of the universal crossovers, employing the language of
the Anderson model, parallels the procedure used with the Kondo model,
and is again compared with success to the numerical data.
A field-theoretic derivation T-matrix in the Kondo problem is given in the
appendix.
In the course of the paper, some previously unaddressed technical issues will
also be discussed, such as the absence of double-counting problems in
renormalized perturbation theory.

\section{The pseudogap Kondo problem: Summary of previous results}
\label{model}

\subsection{Kondo and Anderson models}

The pseudogap Kondo problem\cite{fradkin} (for a review, see
Ref.~\onlinecite{vojta_review}) originates from the
question of how a magnetic impurity behaves when coupled to a fermionic bath
with a depleted (power-law) single-particle spectrum at the Fermi level.
This question
is of interest to various condensed-matter systems such as $d$-wave
superconductors, graphene, and zero-gap semiconductors
(and is in contrast to $s$-wave superconductors, where there is a hard gap).
The qualitative understanding of the pseudogap Kondo problem is
that Kondo screening, i.e., the quenching of the local degrees of freedom,
is weakened with respect to the metallic case due to the lack of low-energy states
in the bath.
This results in the possibility to have quantum phase transitions controlled by
the strength of the Kondo coupling,
the amount of particle-hole symmetry breaking and the shape of the DOS.
In the following, the DOS will be fixed at the simplified model form
\begin{eqnarray} \rho(\epsilon )=\sum_k
\delta(\epsilon-\epsilon_k)=N_0 |\epsilon|^r \theta(D^2-\epsilon^2),
\end{eqnarray}
which has an algebraic dependence in energy characterized by the
number $r>0$ (the standard metallic Kondo problem corresponds to $r=0$).
Here, $\epsilon_k$ are single-particle energies of the fermionic bath,
$D$ denotes the corresponding half-bandwidth, and $N_0=(r+1)/(2 D^{r+1})$.
We consider the Kondo model with an
antiferromagnetic coupling $J>0$ between the impurity and the bath:
\begin{eqnarray}
H_{\rm K} &=& \sum_{k \sigma} \epsilon_k c^{\dagger}_{k
\sigma}c^{\phantom{\dagger}}_{k \sigma}+J \vec{S} \cdot \sum_{\sigma \sigma'}
c^{\dagger}_\sigma (0)\frac{\vec{\tau}_{\sigma
\sigma'}}{2}c^{\phantom{\dagger}}_{\sigma'}(0)\nonumber \\ &+& E_0\sum_\sigma
c^{\dagger}_\sigma (0) c^{\phantom{\dagger}}_\sigma (0)
\label{HKondo}
\end{eqnarray}
in standard notation.
For $E_0=0$ the model is particle-hole symmetric, and $E_0$ tunes the degree of
asymmetry.

A related model is the pseudogap generalization of the Anderson Hamiltonian,
which reads:
\begin{eqnarray}
H_{\rm A} &=&\sum_{k \sigma} \epsilon_k
c^{\dagger}_{k \sigma}c^{\phantom{\dagger}}_{k \sigma}
+ U d^{\dagger}_\uparrow d^{\phantom{\dagger}}_\uparrow
d^{\dagger}_\downarrow d^{\phantom{\dagger}}_\downarrow
\nonumber\\
&+& \sum_\s \left(
\epsilon_d d^{\dagger}_\sigma d^{\phantom{\dagger}}_\sigma
+ V[d^{\dagger}_\sigma c^{\phantom{\dagger}}_\sigma(0)+H.c.]\right).
\label{HAnderson}
\end{eqnarray}
It is worthwhile noting that the Anderson model is equivalent to the Kondo
model in the limit of frozen charge fluctuations, i.e., single occupancy.
This so-called Kondo limit of the Anderson model, is obtained
via Schrieffer-Wolff transformation\cite{hewson} in the limit of large $U$, and leads to
the effective parameters of the Kondo Hamiltonian~(\ref{HKondo}):
\begin{eqnarray}
J & = & 2V^2\left
(\frac{1}{|\epsilon_d|}+\frac{1}{|U+\epsilon_d|}\right), \\
E_0 & = & 2V^2\left
(\frac{1}{|\epsilon_d|}-\frac{1}{|U+\epsilon_d|}\right).
\end{eqnarray}
The particle-hole symmetric case corresponds thus to the value
$\epsilon_d = -U/2$.

\subsection{Fixed-point structure}

\begin{figure}[!t]
\begin{center}
\includegraphics[width=7.0cm]{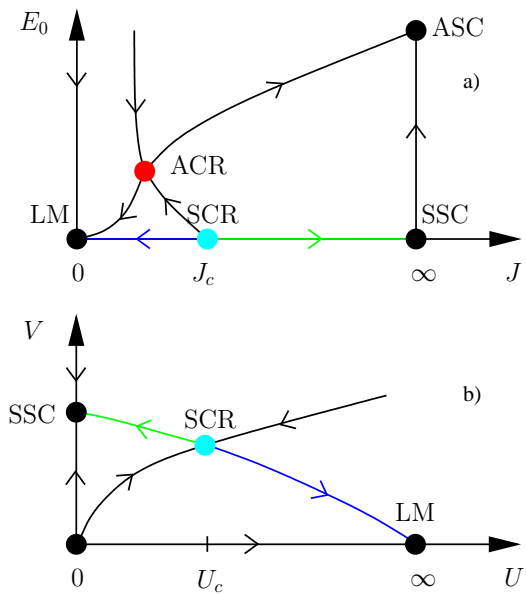}
\end{center}
\caption{(color online)
Generic flow diagrams for
a) the Kondo model (with $r^\star<r<1/2$);
b) the particle-hole symmetric Anderson model (with $r<1/2$). In both cases, the
universal crossovers (depicted as colored lines) will be computed. See the main
text for a description of the nomenclature used in this figure.
}
\label{flow}
\end{figure}

One of the main results of the extensive numerical studies performed by
Gonzalez-Buxton and Ingersent\cite{ingersent} was the generic phase diagram
of the pseudogap Kondo model
(as well as of the pseudogap Anderson model, belonging to the same universality class).
A subset of the results, for an intermediate value of the bath exponent $r$,
is displayed in Fig.~\ref{flow}a as a function of Kondo coupling $J$ and
particle-hole asymmetry $E_0$.
At particle-hole symmetry, $E_0=0$, a quantum critical point (denoted by SCR)
separates the local-moment (LM) phase from the strong-coupling (SSC) regime.
The SCR fixed point
is expelled from the decoupled LM point as soon as $r$ is non zero, but
collapses with the SSC point when $r$ reaches the value 1/2. Away from
particle-hole symmetry, the transition is controlled by SCR only in the range
$0<r<r^\star\simeq0.375$, and for larger values of $r$, a second critical point
(denoted ACR because of its intrinsic asymmetric nature) drives the transition.
Fig.~\ref{flow}a thus corresponds to the intermediate regime $r^\star<r<1/2$ where
both critical points coexist.
Recently, a general understanding of this quite complex phase diagram has been
reached on an analytical level,\cite{fritz1} thanks to the identification of the effective
theories which can be used in the various regimes of the parameter $r$.

For small values of $r$, the Kondo model is the correct effective model to
describe the phase transition between the local-moment and the strong-coupling phases.
Indeed, the critical SCR point at $J=J_c$ moves towards the LM fixed point as
$r$ goes to zero, and the quantum phase transition at particle-hole symmetry
can be controlled by (renormalized) perturbation theory in $J$ within the Kondo
Hamiltonian~(\ref{HKondo}), as discussed in previous works.\cite{fradkin, fritz1}
It is important to point out that this expansion corresponds to the situation
of a {\it lower} critical dimension, where the critical fixed point is located
adjacent to the trivial fixed point around which the expansion is carried out (here
the ``ordered'' LM phase), in analogy to the renormalization flow of the non-linear
sigma model above two dimensions.\cite{zinn}
In principle, we would like to
compute {\it any} physical quantity along the crossover from SCR to LM using the RG
technique -- this will be done in Sec.~\ref{kondo}.

On the other hand, when $r$ approaches the value 1/2, the critical SCR point
moves to strong coupling and the Kondo Hamiltonian is no longer amenable to
a perturbative treatment.
In this case, an expansion around the SSC fixed point is more natural\cite{fritz1},
and the particle-hole symmetric Anderson model~(\ref{HAnderson}) provides the adequate
effective theory, with perturbation theory in $U$ now being controlled in
the small expansion parameter $\epsilon=1-2r$.
Again, the flow diagram in Fig.~\ref{flow}b reflects a situation of {\it lower}
critical dimension.
%, as the transition is controlled by the irrelevant variable $U$.
This full crossover will be calculated in Sec.~\ref{anderson}.
It is important to note that -- within a perturbative framework -- we cannot
describe crossovers associated to a runaway flow of the coupling constant.
Therefore, the crossovers  SCR $\rightarrow$ SSC for small
$r$ and SCR $\rightarrow$ LM for small $(1/2-r)$ small are inaccessible
to analytical calculations.

In the particle-hole asymmetric case, not to be addressed in this paper,
the critical fixed point ACR for $r>r^\ast$ can be accessed using the
perturbative approach of Ref.~\onlinecite{fritz2}, whereas for $r<r^\ast$
multicritical behavior obtains, controlled by the SCR fixed point.

\subsection{Observables}
\label{obs}
To establish some notation, we will introduce the observables that we will
consider in this work, and summarize some results previously obtained
in Ref.~\onlinecite{fritz1}.

\subsubsection{Impurity susceptibility}

Magnetic susceptibilities are generically obtained by coupling an external field
to both the bulk electronic degrees of freedom and the impurity part:
\begin{eqnarray}
H_\mr{field} = \int dx \, \vec{h}_{\rm u}(x) \cdot
\sum_{\s\s'}c^{\dagger}_\sigma(x) \frac{\vec{\tau}_{\sigma \sigma'}}{2}
c^{\phantom{\dagger}}_{\sigma'}(x) + \vec{h}_{\rm imp} \cdot \vec{S}
\end{eqnarray}
where $\vec{S}$ is the impurity spin [it corresponds to
$ \sum_{\s\s'} d^{\dagger}_\sigma \frac{\vec{\tau}_{\sigma \sigma'}}{2}
d^{\phantom{\dagger}}_{\sigma'}$ in the case of the Anderson model].
The bulk field $h_{\rm u}$ varies slowly as function of the space coordinate, and
$h_{\rm imp}$ is the magnetic field at the location of the impurity. With these
definitions, a spatially uniform field applied to the whole system corresponds
to $h_u=h_{\rm imp}=h$. Response functions can be defined from second
derivatives of the thermodynamic potential, $\Omega=-T \textrm{ln} Z$, with respect to
both $h_u$ and $h_{\rm imp}$. Thus $\chi_{u,u}$ measures the bulk response
to a field applied to the bulk only, $\chi_{\rm imp,imp}$ is the impurity response
function to a field which is applied to the impurity only, and $\chi_{\rm u,imp}$
is the crossed response of the bulk to an impurity field.

Following standard practice, we define the impurity contribution to
the total susceptibility (associated to a spatially uniform magnetic field
applied both to the bulk and to the impurity) as
\begin{eqnarray}
\Cimp(T)=\chi_{\rm imp,imp}+2 \chi_{\rm u,imp}+(\chi_{\rm u,u}-\chi^{bulk}_{\rm u,u}),
\end{eqnarray}
where $\chi^{bulk}_{u,u}$ is the susceptibility of the bulk system in absence
of the impurity.

For an impurity spin of size $S=1/2$ we expect the spin to remain
unscreened in the whole LM phase:
\begin{equation}
\Cimp(T\to0,J<J_c)=\frac{1}{4T}.
\end{equation}
(Here and in the following we will use units such that $\hbar=k_B=1$.)

At criticality $\Cimp$ does not acquire an anomalous dimension (in contrast 
to $\Cloc$, see below), because it is the response function associated 
with a conserved quantity, the total spin of the system $S_{\rm tot}$.
In fact, one finds a renormalized Curie law,
\begin{equation}
\Cimp(T\to0,J=J_c)=\frac{C_{\rm imp}(r)}{T} \,,
\end{equation}
which can be interpreted as the Curie response of a fractional
spin, since the universal number $C_{\rm imp}(r)<1/4$.

\begin{figure}[!t]
\begin{center}
\includegraphics[width=7.0cm]{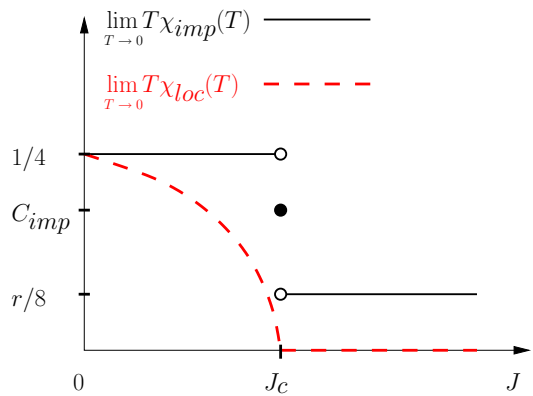}
\end{center}
\caption{(color online)
Low-temperature limit of $T\Cimp$ (continuous upper line) and
$T\Cloc$ (dashed lower line) as a function of the Kondo coupling $J$.
}
\label{chi}
\end{figure}

Finally, the strong-coupling fixed point SSC provides a modified (albeit trivial)
Curie response:
\begin{equation}
\Cimp(T\to0,J>J_c)=\frac{r}{8T}.
\end{equation}
All these results are summarized in Fig.~\ref{chi}.

The phase transition occurring at particle-hole symmetry is
described by an interacting fixed point and thus obeys hyperscaling.
The static susceptibility must therefore satisfy the relation
\begin{eqnarray}
\Cimp(T,J)=\frac{1}{T}
\Phi_{\rm imp} \left(\frac{T}{T^\star} \right),
\label{Phiimp}
\end{eqnarray}
where $T^\star\propto |J-J_c|^\nu$ is the energy scale above which quantum critical behavior
is observed, and $\nu$ is the associated correlation ``length'' exponent.
In this formula $\Phi_{\rm imp}$ is a universal crossover function. Such a
scaling form (and the other ones written below) will be derived in Sec.~\ref{kondo}.

\subsubsection{Local susceptibility}

The local susceptibility is given by
\begin{eqnarray}
\Cloc(T)=\chi_{\rm imp,imp}\,,
\end{eqnarray}
which is equivalent to the zero-frequency impurity spin autocorrelation
function. In the LM phase, one has a Curie law $\Cloc\propto 1/T$
at low temperature, with a prefactor {\em not} pinned to the value $1/4$,
as, contrarily to $\Cimp$,
the local susceptibility is not associated to a conserved quantity.
We can then define a ``magnetization'' $m_{\rm imp}$ as:
\begin{equation}
\lim_{T\to 0} T \Cloc(T,J<J_c)=m^2_{\rm imp}(J)
\label{mimp}
\end{equation}
which turns out to be a suitable order parameter for the quantum phase
transition,\cite{ingersent} as this quantity vanishes continuously when
$J\to J_c$, see Fig.~\ref{chi}.
Physically, the moment of size 1/2 is spatially smeared, and $m_{\rm imp}$ measures
its fraction which remains localized at the impurity site.

At criticality, $m^2_{\rm imp}=0$, and the local susceptibility follows a
non-trivial power law,
\begin{equation}
\Cloc(T\to0,J=J_c)\propto\frac{1}{T^{1-\eta_\chi}}\,,
\label{chiloceta}
\end{equation}
defining the universal anomalous exponent $\eta_\chi$
which controls the anomalous decay of the two-point correlations of the impurity
spin (see below). One obtains a similar behavior for $\Cloc$ at $J>J_c$, although
$\eta_\chi$ takes then a trivial value.

Because of the hyperscaling property, the static local susceptibility follows
\begin{eqnarray}
\Cloc(T,J)=\frac{\mathcal{B}_1}{T^{1-\eta_\chi}}
\Phi_1 \left(\frac{T}{T^\star} \right),
\label{scale_static}
\end{eqnarray}
where $\Phi_{1}$ is a universal crossover function, whereas $\mathcal{B}_{1}$
is a non-universal prefactor (this is related to the fact that $\Cloc$ is
associated to a non-conserved quantity).

Hyperscaling also implies a scaling in $\omega/T$ for the dynamic quantities.
The dynamic local susceptibility can be defined from the imaginary-time
correlation function:
$\Cloc(\tau,T,J) = \big< \vec{S}(\tau) \cdot \vec{S}(0)\big>$.
In particular, the spectral density of its Fourier transform obeys the following
scaling form
\begin{eqnarray}
\chi''_{\rm loc}(\omega,T,J)=\frac{\mathcal{B}_2}{\omega^{1-\eta_\chi}}
\Phi_2\left(\frac{\omega}{T},\frac{T}{T^\star} \right),
\label{scale_dynamic}
\end{eqnarray}
with again $\Phi_{2}$ a universal crossover function, and $\mathcal{B}_{2}$
a non-universal number.

\subsubsection{Impurity entropy}
Quantum critical points in impurity models show in general
a finite residual entropy (this is to be contrasted with bulk quantum critical
points, where the entropy has to vanish at zero temperature). The flow
of the coupling constant in the vicinity of the critical point will then imply
a crossover of the entropy, in a similar fashion to the behavior of the Curie
constant associated to the impurity susceptibility.

\subsubsection{T-matrix}

An important quantity in quantum impurity models is the conduction electron T-matrix,
which describes the scattering of conduction electrons off the impurity. For the
Anderson model, the T-matrix is
given by the relation $\T(\omega)=V^2 G_d (\omega)$, with $G_d(\omega)$ the full
$d$-level propagator.
For the Kondo model, one defines a propagator $G_T$ of
the composite operator $T_{\sigma}=\vec{\textrm{S}} \cdot \sum_{\sigma'}
c^{\phantom{\dagger}}_{\sigma'}(0) \vec{\tau}_{\sigma \sigma'}$ such that the
T-matrix is given by $\T(\w)=J^2 G_T(\omega)$
(see App.~\ref{apptmatrix} for a derivation of this
formula, as well as Refs.~\onlinecite{kircan,costi,zarand2,zarand3}).

The hyperscaling ansatz gives the general expression
\begin{eqnarray}
\T(\omega,T,J)=\mathcal{B}_\T \; \omega^{r-\eta_{\T}}
\Phi_\T \left(\frac{\omega}{T},\frac{T}{T^\star} \right) \,,
\label{hyper_T}
\end{eqnarray}
where an anomalous exponent $\eta_\T$ was introduced.
One has in fact the {\it exact} result\cite{kircan} $\eta_\T=2r$,
so that the critical T-matrix behaves as
\begin{eqnarray}
\T(\omega,T=0,J=J_c)\propto \frac{1}{\omega^{r}} \,.
\end{eqnarray}

In the remainder of the paper, we intend to perturbatively calculate
all of the scaling functions defined above.

\section{Crossover functions near $r=0$:
Weak-coupling regime of the Kondo model}
\label{kondo}

As discussed in Ref.~\onlinecite{fritz1}, the Kondo model is the correct 
starting point to describe the phase transition between the LM and SSC
phases when the parameter $r$ is small.
In analogy to the theory of critical phenomena, $r=0$ constitutes a lower critical
dimension of the problem, and
the RG for the pseudogap Kondo model is analogous to the RG in the non-linear
$\sigma$-model. The RG equations have first been derived by Withoff and
Fradkin,\cite{fradkin} who pointed out the possibility to have a quantum phase
transition driven by the strength of the Kondo coupling.

One can switch to the dimensionless renormalized coupling $j\equiv N_0 Z_j^{-1} Z_f \mu^r J$,
where $\mu$ is an arbitrary renormalization energy scale,
and $Z_j$ ($Z_f$) are the vertex (field) renormalization factors.
To one-loop order, the RG equation reads\cite{fradkin,si1,zarand1,kircan}
\begin{eqnarray}
\beta(j)\equiv \frac{\textrm{d}j}{\textrm{dln}\mu}=r j-j^2 \,,
\label{scaling}
\end{eqnarray}
and the renormalization factors evaluate to
\begin{equation}
Z_j = 1+\frac j r \,,~~ Z_f = 1\, .
\end{equation}
The beta function (\ref{scaling}) yields an infrared unstable fixed point at
\begin{eqnarray}
j_c =r+\mathcal{O}(r^2)
\end{eqnarray}
which controls the transition between LM and SSC.
Because $j$ is dimensionless, we will prefer to write the differential 
equation (\ref{scaling}) w.r.t. a dimensionless parameter $\lambda$, 
and integrate it between the values $j$ for $\lambda=1$ (corresponding to
the initial value of the renormalized coupling at the renormalization 
scale $\mu$) and $j(\lambda)$ at the lower observation energy scale 
$\lambda\mu$, with $\lambda<1$.
The solution reads:
\begin{eqnarray}
j(\lambda)
& = & \frac{r}{1+ [(r-j)/j] \; \lambda^{-r}} \nonumber \\
& = & \frac{r}{1+ \mr{sgn}(r-j) \; |\lambda\mu/T^\star|^{-r}}
\label{running_j} \\
\mr{with} \;\; T^\star & \equiv & \mu \left|\frac{r-j}{j}\right|^{1/r} \,.
\end{eqnarray}
We have introduced the energy scale $T^\star$, which vanishes at the
quantum critical point, and generally sets the crossover scale between the
various fixed points.
We notice first that the running coupling obeys on the one hand $j(\lambda) = r$
if $j=r$ (fixed-point behavior), while $j(\lambda) \propto \lambda^r$ if $j<r$
and $\lambda\mu\ll T^\star$ (flow to the weak-coupling LM fixed point).
In this case, the running coupling
constant is therefore always perturbatively small (when $r\ll1$), and one should
be able to follow the full crossover between the critical point and the LM phase.
On the other hand, for $j>r$ the above expression shows a pole, that signals that
the crossover between the critical point and the SSC phase occurs
through runaway flow from weak coupling, invalidating the perturbative expansion.
This regime is analogous to the metallic Kondo problem, where the perturbative
expansion breaks down at the scale $T_K$, the Kondo
temperature,\cite{hewson} which can be here associated to $T^\star$ for $j>r$
(one can check that as $r$ vanishes, the usual one-loop expression for the Kondo
temperature is recovered from $T^\star$).
In the following, we will focus on the perturbatively accessible crossover
between SCR and LM (corresponding to $j<r$), while the crossover from SCR to SSC
will be tackled through the mapping onto the Anderson model discussed in
Sec.~\ref{anderson}.

\subsection{Regular observables}
\label{sec:regular}

Within this section we will calculate the crossover functions for observables
which do not show anomalous power laws.
Technically, this means
that the renormalization $Z$-factors for these observables are strictly unity and
perturbative corrections do not change the power-law behavior of the tree level
result, but imply a renormalized universal amplitude.
Usually such observables are associated with conserved quantities,\cite{buragohain}
such as the total spin of the system in the case of the impurity susceptibility.

\subsubsection{Impurity susceptibility}

\begin{figure}[!t]
\begin{center}
\includegraphics[width=8.5cm]{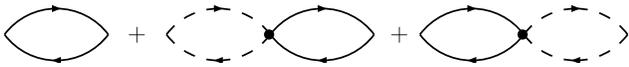}
\end{center}
\caption{
Diagrammatic expansion for the impurity susceptibility of the Kondo
model, with the tree level contribution (of $\chi_{\rm imp,imp}$ type) plus the order $j$
perturbative correction (of $\chi_{\rm u,imp}$ type). Here full lines denote
the Abrikosov fermion propagator, while dashed lines represent the bulk fermions.
Each vertex, labelled by a dot, yields the value $J=N_0^{-1} Z_j Z_f^{-1} \mu^{-r} j$.}
\label{chiim}
\vspace{0.5cm}
\end{figure}

One can calculate the crossover function for the impurity
susceptibility from the perturbative result derived in Ref.~\onlinecite{kircan}. The
diagrammatics is done introducing a constrained Abrikosov fermion
$f^\dagger_\s$ to represent the impurity spin
$\vec{S} = \sum_{\s\s'} f^\dagger_\s \frac{\vec{\tau}_{\s\s'}}{2} f^{\phantom{\dagger}}_{\s'}$,
with the constraint $\sum_\s f_\s^\dagger f_\s = 1$.
To lowest order in $j$, the impurity susceptibility is given in
Fig.~\ref{chiim}.
The correction of order $j$ takes the form
\begin{eqnarray}
\label{chiloccor}
\chi_{\rm u,imp}=-j \mu^{-r} \frac{1}{4 T^2}\int^D_{-D} \textrm{d}\epsilon
|\epsilon|^r \frac{\textrm{cosh}^{-2}\left(\frac{\epsilon}{2 T}\right)}{4},
\end{eqnarray}
which, when evaluated in the limit of infinite UV cutoff $D$ and at lowest order
in $r$, gives
\begin{eqnarray}
\Cimp=\frac{1}{4T}\left[1-\left|\frac{T}{\mu}\right|^r j\right].
\label{pert}
\end{eqnarray}
using the tree-level result $\Cimp=1/(4T)$ at $j=0$.
An intuitive way to obtain the full crossover from SCR to LM is to set $\mu=T$
in the previous expression and to plug in the running coupling $j(\lambda)$,
Eq.~(\ref{running_j}), taken at a scale $\lambda=T/\mu$.
However, we would like
to derive this result rigorously in the simplest case of $\Cimp$, in view of the more
complicated physical quantities we will consider later on.

In the following we resort to the concepts of renormalization,\cite{zinn,peskin}
familiar in the field-theoretic context.
We start by noting that the {\it bare} impurity susceptibility $\Cimp^B$ explicitly depends
on the high-energy cutoff $D$ but should be insensitive to the
arbitrary energy scale $\mu$,
while conversely the {\it renormalized} $\Cimp$ is independent of $D$ (which
is ultimately taken to infinity in the renormalized theory).
The two are in general related through
\begin{equation}
\Cimp^B(T,j,D) = Z_{\rm imp} \Cimp(T,j,\mu)
\end{equation}
where the renormalization factor $Z_{\rm imp}=1$ here, because one is dealing with a
conserved quantity free of multiplicative renormalizations.
We therefore have
\begin{eqnarray}
\mu \frac{\mr{d}}{\mr{d}\mu} \Cimp^B(T,j,D) & = & 0\\
\Rightarrow
\left[ \mu \frac{\partial }{\partial \mu} + \beta(j) \frac{\partial}{\partial j}
\right] \Cimp(T,j,\mu) & = & 0
\label{callan_imp}
\end{eqnarray}
with $\beta(j) = \textrm{d}j/\textrm{dln}\mu$.
This is a special case of a Callan-Symanzik (CS) equation, which reflects the
invariance of the renormalized theory with respect to the high-energy cutoff,
and therefore is a strong statement of universality.
It is solved by:
\begin{equation}
\Cimp(T,j,\mu) = F\left(\ln \mu - \int^j \frac{d j'}{\beta(j')}\right)
\end{equation}
where $F$ is an arbitrary function.
From the $\beta$ function~(\ref{scaling}), one has
$\ln \lambda = \int^{j(\lambda)}_j dj'/\beta(j')$, which inserted in the
previous equation gives:
\begin{eqnarray}
\Cimp(T,j,\mu) & = & F\left(\ln (\lambda\mu) - \int^{j(\lambda)} \frac{d
j'}{\beta(j')}\right) \\
\Rightarrow \Cimp(T,j,\mu) & = & \Cimp(T,j(\lambda),\lambda\mu)
\end{eqnarray}
We finally use the fact that $T$, $\mu$, and $1/\Cimp$ have dimensions of
energy, to obtain:
\begin{equation}
\Cimp(T,j,\mu) = \frac{1}{\lambda}
\Cimp\left(\frac{T}{\lambda},j(\lambda),\mu\right) \,.
\label{CS_imp}
\end{equation}
This type of equation is the starting point for the calculation of the
crossover functions defined in Sec.~\ref{obs}. Because $\mu$ appears
identically on both sides of~(\ref{CS_imp}), we will drop it in the following.
We proceed by inserting the perturbative result~(\ref{pert}) in the
r.h.s. of the previous equation:
\begin{equation}
\Cimp(T,j) = \frac{1}{4T}
\left[
1-\left|\frac{T}{\lambda\mu}\right|^r j(\lambda) + {\mathcal O}(j^2) + \ldots
\right]\,.
\label{chi_lam}
\end{equation}

We now discuss the choice of the renormalization parameter $\lambda$ which
is crucial for all subsequent computations.
Generally, an equation like (\ref{chi_lam}) is valid for any $\lambda$
(provided that all terms of the expansion are included!),
however, for the application of a perturbative scheme we have to
{\em require} the higher-order corrections to be small.
As these in general have a singular structure, there is usually a
{\em unique} choice of $\lambda$ (up to a perturbative re-definition)
to obtain a controlled perturbative expression.
In the present case of $\Cimp$, $j^2$ corrections (displayed explicitly in
Sec.~\ref{sec:chiloc} below) display $\ln(T/\lambda\mu)$ singularities.
This fixes $\lambda$ to $\lambda=T/\mu$ to allow perturbation theory
to be valid, and this yields:
\begin{equation}
\Cimp(T,j) = \frac{1}{4T}
\left[1-j(T/\mu)\right] \,.
\end{equation}
Using the running coupling $j(\lambda)$, Eq.~(\ref{running_j}),
we finally get:
\begin{eqnarray}
%T \Cimp=\frac{1}{4}\left[1-\frac{r}{1+\left(\frac{T}{T^\star}\right)^{-r}}\right]
\nonumber
T \Cimp & =& \Phi_{\rm imp}\left(\frac{T}{T^\star}\right) \,, \\
\Phi_{\rm imp}(x) & = & \frac{1}{4}\left[1-\frac{r}{1+x^{-r}}\right]
\label{tchiimp}
\end{eqnarray}
which is the desired scaling form (\ref{Phiimp}).
Formula~(\ref{tchiimp}) interpolates smoothly between the free-spin value $1/4$
at low temperature ($T\ll T^\star$) and the non-trivial fractional spin value
$(1/4)(1-r)$ associated to the quantum critical high-temperature regime
($T\gg T^\star$).

\begin{figure}[!t]
\begin{center}
\includegraphics[width=7.8cm]{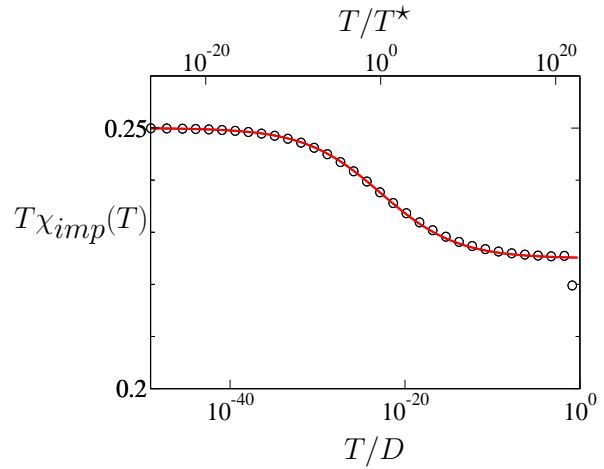}
\end{center}
\caption{(color online)
Impurity susceptibility $T\Cimp(T)$ at $r=0.1$ given in Eq.~(\ref{tchiimp}) (solid),
capturing the crossover between SCR at high temperature and LM at low temperature.
Also shown are NRG data (dots) for $T\Cimp$, obtained for
a pseudogap Anderson model\cite{nrgpara} with $U/D=1$, $N_0V^2/D=0.05$.
The crossover scale $T^\ast$ has been fitted: $T^\star/D \simeq 1.7\times10^{-23}$.
[The reason for the very slow crossover from SCR to LM is the large value of the
correlation length exponent, $\nu=1/r + \mathcal{O}(1) \approx 10$.] 
Note also that deviations from scaling behavior occur when temperature is
comparable to the bandwidth $D$, as seen by a kink in the last data point.}
\label{TChi_r0.1}
\vspace{0.5cm}
\end{figure}

\begin{figure}[!t]
\begin{center}
\includegraphics[width=8.2cm]{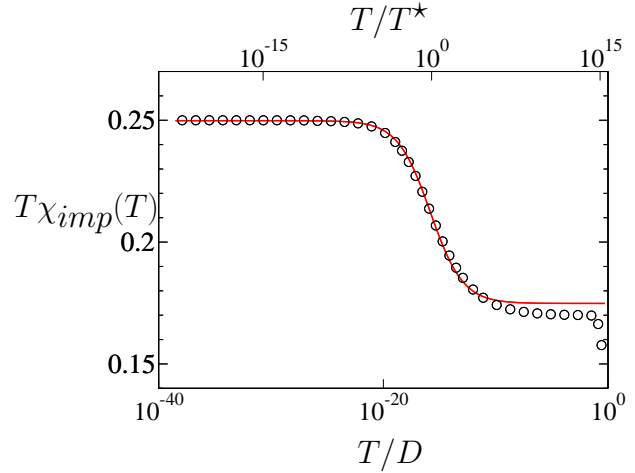}
\end{center}
\caption{(color online)
Impurity susceptibility $T\Cimp(T)$, Eq.~(\ref{tchiimp}), now for $r=0.3$,
together with NRG data for $U/D=1$, $N_0V^2/D=0.25$, resulting in
$T^\star = 2\times10^{-16}$.
As discussed in the text, the agreement between the analytical curve and the NRG data
is better at low temperature, where the running $j(\lambda)$ has flown towards
weaker coupling.
}
\label{TChi_r0.3}
\end{figure}

Expression~(\ref{tchiimp}) can be compared to NRG simulations (see Figs. \ref{TChi_r0.1}
and \ref{TChi_r0.3}), with $T^\star$ as the unique fitting parameter. For $r=0.1$ the agreement
is seen to be excellent, over 40 orders of magnitude! We note that for the larger value
$r=0.3$, the fit deteriorates, but only at high temperature. This is because the
critical point does not lie anymore in the perturbative regime (a two-loop
calculation may improve this result).
At low temperature, however, the agreement becomes better, a fact that we
interpret from the flow of the running coupling constant towards weak coupling.

\subsubsection{Impurity entropy}
\label{entropy}

The same calculation in the case of the impurity entropy shows the perturbative
result:\cite{kircan}
\begin{equation}
\Simp=\ln{2}\left[1+\frac{3\pi^2}{8}j^3\left|\frac{T}{\mu}\right|^{3r}\right].
\end{equation}
Application of the CS equation with $\lambda=T/\mu$ gives:
\begin{eqnarray}
\Simp&=&\ln{2}\left[1+\frac{3\pi^2}{8}j^3\left(T/\mu\right)\right] \nonumber \\
&=&\ln{2}\left[1 +\frac{3\pi^2}{8}
\left(\frac{r}{1+(T/T^\star)^{-r}}\right)^3\right].
\end{eqnarray}
Since the observed variations in the impurity entropy are actually quite
small for the pseudogap Kondo model,\cite{fritz1} we will not try to compare this
quantity with the numerical results.

\subsection{Local susceptibility}
\label{sec:chiloc}

In this section we will present a detailed calculation of the static and
dynamic local susceptibility, $\chi_{\rm loc}$, within the weak-coupling
expansion of the Kondo model using the CS equations.
In contrast to the quantities of Sec.~\ref{sec:regular}, $\Cloc$
does display singular behavior with anomalous power laws.

The starting point is the calculation of the
renormalization factor $Z_{\chi}$, since we now
deal with a quantity which is not conserved.
The corresponding diagrammatic expressions are in Refs.~\onlinecite{si1,zarand1}, and yield
[this result will be re-derived below, see Eq.~(\ref{line2})]:
\begin{equation}
Z_\chi = 1 + \frac{j^2}{2r} + O(j^3)
\label{Zchi}
\end{equation}
As before, one can relate the bare susceptibility $\Cloc^B$ to the
renormalized one $\Cloc$ \footnote{If one wants to compare this correlation
function to standard literature in field theory, the identification of
$\chi$ and the vertex $\Gamma^{0,2}$ can be made.} by:
\begin{eqnarray}
\Cloc^B(\omega,T,J,D)=Z_{\chi} \Cloc(\omega,T,j,\mu)
\end{eqnarray}
The next step is to derive the bare quantity with respect to the
renormalization scale $\mu$, which after some straightforward algebra
leads to the Callan-Symanzik equation for the renormalized local susceptibility
\begin{eqnarray}
\left[\mu \frac{\partial}{\partial \mu}+\beta(j) \frac{\partial}{\partial j}+\eta_\chi(j)
\right] \Cloc(\omega,T,j,\mu)=0
\label{callan_loc}
\end{eqnarray}
with the anomalous exponent
\begin{equation}
\eta_\chi(j)=\beta(j)\frac{\partial \ln{Z}_{\chi}}{\partial j} = j^2 + O(j^3) \,.
\end{equation}
[Because $Z_{\rm imp}=1$, the CS equation~(\ref{callan_imp}) for $\Cimp$ lacks such an
extra term.]
The main relation in the following calculations
is the integrated Callan-Symanzik equation for the local susceptibility, which
is given by
\begin{eqnarray}
\Cloc(\omega,T,j) & = & \frac{1}{\lambda} f_\chi(\lambda)
\Cloc\left(\frac{\omega}{\lambda},\frac{T}{\lambda},j(\lambda)\right)
\label{CS_loc}
\end{eqnarray}
The factor $f_\chi(\lambda)$ is at the origin of all anomalous power-law
behaviors as we will see shortly, and reads:
\begin{eqnarray}
\label{fchi}
f_\chi(\lambda) & = &
\exp\left(\int_j^{j(\lambda)} \!\!\!\!
dj'\frac{\eta_\chi(j')}{\beta(j')}\right) \\
\nonumber
& = & \left(\frac{r-j}{r}\right)^r \exp(j-r)
\exp\left(\frac{r}{1+|\lambda \mu/T^\star|^r}\right)\\
& & \times \left[1+\left|\frac{\lambda\mu}{T^\star}\right|^r\right]^r
\nonumber
\end{eqnarray}
At the quantum critical point, $j=r$, the above expression reduces to
$f_\chi(\lambda) = \lambda^{r^2}$.

Note that the choice $\lambda=\w/\mu$ in~(\ref{CS_loc}) proves rigorously the
$\w/T$ scaling property stated in Eq.~(\ref{scale_dynamic}).
It is important to understand that such scaling behavior generally applies only below
the upper critical dimension, but as the particle-hole symmetric Kondo problem
only possesses {\it two} lower critical dimensions, the scaling ansatz is
fulfilled for all values of $0<r<1/2$ (outside this range the critical point
ceases to exist).
In the particle-hole asymmetric case, hyperscaling is violated when $r>1$ due
to the presence of a dangerously irrelevant coupling.\cite{fritz1}
In such a case, the observable vanishes in the naive scaling limit
(because the fixed point value of the coupling is zero), and
the leading behavior is exposed only upon considering corrections
to scaling (which do not follow hyperscaling laws).

\subsubsection{Impurity magnetization}

\begin{figure}[!t]
\begin{center}
\includegraphics[width=8.0cm]{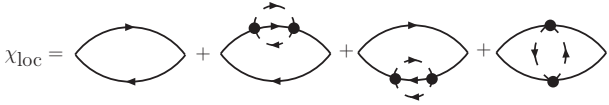}
\end{center}
\caption{
Diagrams contributing to the local susceptibility $\Cloc$ of the
Kondo model up to order $j^2$, with conventions given in 
Fig.~\protect\ref{chiim}. For the second and third graphs above, 
the self-energy-like contributions require a mass substraction.}
\label{chilocrsmall}
\vspace{0.5cm}
\end{figure}

A quantity, which turns out to be a good order parameter for the transition is
the impurity magnetization, which is related to the low-temperature limit
of the zero-frequency local susceptibility by Eq.~(\ref{mimp}).
We start, as previously, by a direct perturbative calculation of
$\Cloc(T,j) \equiv \Cloc(\w=0,T,j)$.
To second order in $J$ the diagrams shown in Fig.~\ref{chilocrsmall} evaluate to:
\begin{eqnarray}
\nonumber
T \Cloc(T,j) & = & Z_{\chi}^{-1}\left(
1 + \frac{j^2}{2}\left|\frac{T}{\mu}\right|^{2r}
\int_{-\infty}^{\infty} \!\!\!\! dx
\int_{-\infty}^{\infty} \!\!\!\! dy |x|^r|y|^r \right. \\
\nonumber
& & \hspace{-2.2cm} \left. \frac{e^{-x}}{\left(e^{-x}+1\right)\left(e^{-y}+1\right)
\left(x-y\right)^2} \left[2 \frac{e^{x-y}-1}{x-y}-1-e^{x-y}\right] \right).
\end{eqnarray}
Note that the quantity in brackets is the perturbative expression for the
{\em bare} susceptibility, where the bare coupling $J$ has been expressed in
terms of the renormalized $j$, and the limit of infinite bandwidth has been
taken. Further, the additional renormalization factors $Z_j$ and $Z_f$ do not
appear as they can be approximated by unity to the order we are working here.
Evaluating the integrals yields:
\begin{eqnarray}
\nonumber
T \Cloc(T,j)
& = &Z_\chi^{-1} \left(
1 + j^2 \left|\frac{T}{\mu}\right|^{2r}
\left(\frac{1}{2r}-\ln\frac{e^{\gamma}}{2\pi}\right)
\right)\\
& = & 1 - j^2 \ln\frac{e^{\gamma}\mu}{2\pi T}
\label{line2}
\end{eqnarray}
where $\gamma = 0.5772...$ is Euler's number.

We can now apply the CS equation~(\ref{CS_loc}) at zero frequency:
\begin{eqnarray}
\label{chipert}
\Cloc(T,j) & = & \frac{1}{\lambda} f_\chi(\lambda)
\Cloc\left(\frac{T}{\lambda},j(\lambda)\right) \\
\nonumber
& = & \frac{1}{\lambda} f_\chi(\lambda)
\left(1 - j^2\ln\frac{e^{\gamma}\lambda\mu}{2\pi T}
\right)
\end{eqnarray}
We see again that the particular choice $\lambda=T/\mu$ allows to control the
perturbative expansion.
Putting together~(\ref{fchi}) and (\ref{chipert}), we therefore arrive to the
expression:
\begin{eqnarray}
\nonumber
\Cloc(T,j)&=& \frac{\mcal{C}_1}{T}
\left[1+\left(\frac{T}{T^\star}\right)^r\right]^r
\exp\left(\frac{r}{1+|T/T^\star|^r}\right)\\
\mr{with} \; \; \mcal{C}_1 &=& \left(\frac{r-j}{r}\right)^r
\left(1-j^2\ln\frac{e^{\gamma}}{2\pi}\right)
\label{chilocfinal}
\end{eqnarray}
which is the scaling form~(\ref{scale_static}), with the anomalous exponent
\begin{equation}
\eta_\chi = r^2.
\end{equation}

\begin{figure}[!t]
\begin{center}
\includegraphics[width=8.4cm]{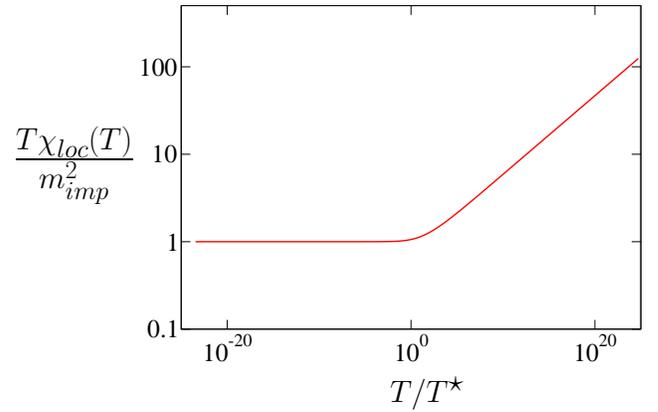}
\end{center}
\caption{(color online)
Rescaled static local susceptibility $T\Cloc(T)$ for $r=0.3$ according to
formula~(\ref{chilocfinal}). Although not performed here, a comparison to NRG data
could be possible.}
\label{TChiloc}
\vspace{0.5cm}
\end{figure}

In the limit $T\ll T^\star$, we find, as expected, that the magnetization vanishes
as a power law at the quantum critical point:
\begin{eqnarray}
T\Cloc(T,j) & = & m_{\rm imp}^2 \\
\mr{with} \;\; m_{\rm imp} & \propto & (j_c-j)^{r/2}
\end{eqnarray}
However, when $T\gg T^\star$, critical behavior with an anomalous temperature
dependence obtains:
\begin{equation}
\Cloc(T,j) \propto \frac{1}{T^{1-r^2}}
\end{equation}
as anticipated in Eq.~(\ref{chiloceta}). The complete crossover function,
defined in equation~(\ref{chilocfinal}), is shown in figure~\ref{TChiloc}.

\subsubsection{Dynamic susceptibility}
\label{secchi}

We are now interested in the imaginary part of the real frequency susceptibility
$\Cloc''(\w,T,j)$. It is actually easier to perform the perturbative
calculation on the imaginary-time axis.
The graphs displayed in Fig.~\ref{chilocrsmall}
give a complicated integral (not shown), which, however, simplifies considerably
by taking a double time derivative:
\begin{eqnarray}
\nonumber
\frac{\partial^2\Cloc}{\partial \tau^2} & = &\frac{j^2}{\tau^2}+ j^2 \mu^{-2r} \int d\epsilon
|\epsilon|^r \int d\epsilon' |\epsilon'|^r
e^{(\epsilon'-\epsilon)\tau}\\
& & \times
\left[\frac{1}{e^{-\beta\epsilon}+1}
\frac{1}{e^{\beta\epsilon'}+1}-\theta(\epsilon)\theta(-\epsilon')\right] \\
& = & j^2 \left(\frac{\pi T}{\sin(\pi T \tau)}\right)^2 \,.
\end{eqnarray}
where $\beta$ above defines the inverse temperature $1/T$.
This expression is readily integrated, using Eq.~(\ref{line2}) to obtain
the integration constant, which gives the simple result:
\begin{eqnarray}
\Cloc(\tau) & = & 1 - j^2
\ln\frac{e^{\gamma}\mu \sin(\pi T \tau)}{\pi T} \,.
\label{chiperttau}
\end{eqnarray}

\begin{figure}[!t]
\begin{center}
\includegraphics[width=8.4cm]{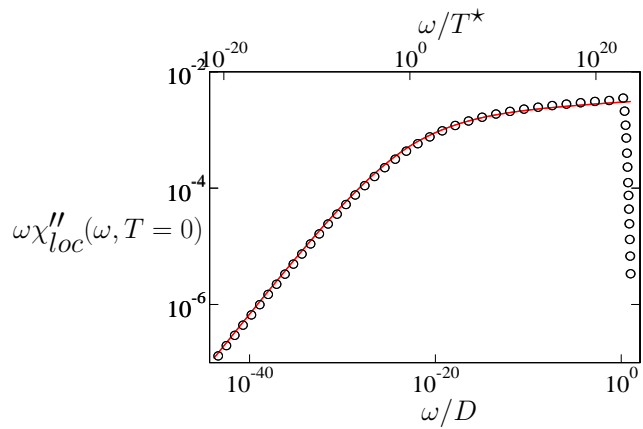}
\end{center}
\caption{(color online)
Rescaled dynamic local susceptibility
$\w\Cloc''(\w,T=0)$ at zero temperature for $r=0.1$, with the same parameters
as in Fig.~\ref{TChi_r0.1}.
NRG data points are the dots, and the solid line
follows formula~(\ref{Chiloc_w}). Since the crossover scale $T^\star$ was already
determined from $\Cimp(T)$, the only fitting parameter here is the
non-universal amplitude $\mcal{C}_2$.}
\label{Chiloc_r0.1}
\vspace{0.5cm}
\end{figure}

Now performing the Fourier transform and the analytical continuation, one finds:
\begin{equation}
\Cloc''(\w) =
\left[1 - j^2 \ln\frac{e^{\gamma}\mu}{2\pi T}\right] \pi \beta \w \delta(\w)
+ \pi \frac{j^2}{\w} \,.
\label{chilocpert}
\end{equation}
The zero-frequency contribution corresponds to the Curie part of the
static susceptibility, see Eq.~(\ref{line2}), while the finite-frequency
term is associated to dissipation.
Let us first focus on the zero-temperature case. Applying the CS
equation~(\ref{CS_loc}), one gets at finite frequency (for readability, we
neglect here the exponential factor which only provides a small renormalization
of the amplitude):
\begin{equation}
\Cloc''(\omega,T=0,j) = \frac{1}{\w} \mcal{C}_2 j(\lambda)^2
\left|\frac{\w}{\lambda\mu}\right|^{2r}
\left[1+\left|\frac{\lambda\mu}{T^\star}\right|^r\right]^r .
\end{equation}
Employing $\lambda=\w/\mu$ (this is again vindicated by requiring
that the next term appearing in perturbation theory should be smaller than the
present leading contribution), we finally obtain:
\begin{equation}
\Cloc''(\omega,T=0,j) = \frac{1}{\w} \mcal{C}_2
\left[\frac{r}{1+|\w/T^\star|^{-r}}\right]^2
\left[1+\left|\frac{\w}{T^\star}\right|^r\right]^r .
\label{Chiloc_w}
\end{equation}
This scaling function crosses over from $\w^{-1+2r}$ behavior at $\w\ll T^\star$ to
$\w^{-1+r^2}$ behavior at $\w\gg T^\star$, as can be seen in
Fig.~\ref{Chiloc_r0.1}.

We have thus obtained the zero-temperature limit of the scaling
function~(\ref{scale_dynamic}), and can now wonder whether this program
can be fulfilled at finite temperature as well. Of course, since the regime
where $\w \gg T$ obeys the zero-temperature scaling form, one is rather
interested in the so-called ``relaxational'' ($\w\ll T$) and
crossover ($\w\sim T$) regimes.
It is easy to see that a direct application of the renormalization equations on
the real frequency axis fails for $\w\ll T$.
Indeed, by the Kramers-Kronig relation, the
static local susceptibility is related to the spectral density by:
\begin{equation}
\Cloc(T) = -\frac{1}{\pi}\int d\w \frac{\Cloc''(\w,T)}{\w}
\end{equation}
At non-zero temperature, the l.h.s. is finite and given by
Eq.~(\ref{chilocfinal}), so that $\chi_{loc}''(\w,T)$, an odd function of $\w$,
has to vanish at low frequency. However, the $1/\w$ divergency shown
in~(\ref{chilocpert}) is in contradiction with the above statement [note
that the application of the CS equation cannot allow to fully remove this
$1/\w$ divergency]. The reason for the failure of this calculation is the
non-commutativity of the renormalization procedure and the analytic continuation
in the limit $\w\ll T$, as was already noted in the context of quantum
antiferromagnets by Sachdev.\cite{sachdev}

We will now show that the application of the CS equation {\it before} the
analytic continuation (namely on the imaginary time axis) does allow to recover
a physically sound result (although the general applicability of this procedure
remains to be benchmarked). Indeed, the perturbative expression~(\ref{chiperttau})
implies the choice $\lambda= \pi T/[\mu\sin(\pi T\tau)]$ of the renormalization
parameter to remove the logarithmic divergency. For simplicity, we assume that
one sits at the critical point ($j=r)$, so that $f_\chi(\lambda) = \lambda^{r^2}$
in equation~(\ref{fchi}).
The CS equation for the $\tau$-dependent susceptibility gives immediately (up to
a small multiplicative correction):
\begin{equation}
\Cloc(\tau,T,j=r) = \left(\frac{\pi T}{\mu\sin(\pi T\tau)}\right)^{r^2}
\end{equation}
This well-defined function can be Fourier-transformed and
analytically continued:\cite{si2,parcollet}
\begin{eqnarray}
\nonumber
\Cloc(\w,T,j=r) & = & \frac{1}{T} \left(\frac{2\pi T}{\mu}\right)^{r^2}
\!\!\frac{1}{\pi}\sin(\pi r^2/2) \Gamma(1-r^2) \\
&\times& \frac{\Gamma(r^2/2+i\w/2\pi T)}{\Gamma(1-r^2/2-i\w/2\pi T)}
\end{eqnarray}
One can then expand this expression with respect to $\w$ and $j=r$, which gives
two {\it different} results according to the order in which the limits are taken:
\begin{eqnarray}
\nonumber
\lim_{j\rightarrow0}\lim_{\w\rightarrow0} \Cloc(\w,T,j=r) &=& \frac{1}{T} + i \frac{1}{j^2}\frac{\w}{\pi T^2} \,, \\
\lim_{\w\rightarrow0}\lim_{j\rightarrow0} \Cloc(\w,T,j=r) &=& \pi \frac{j^2}{\w} \,.
\end{eqnarray}
The first expression is physically correct, but is clearly non-perturbative,
while the second expression simply recovers the singular perturbative
correction.

This explicit calculation has clearly demonstrated the non-commutativity of the
epsilon-expansion and the analytic continuation, but has hinted that progresses can
be made when one proceeds in the correct order, namely by the use of the CS equation on
the imaginary-time perturbative expression, followed by the analytic
continuation. In principle, the computation of $\Cloc''(\w,T,j)$ away from
the critical point can be done along this line, although a numerical analytic
continuation would be required.

\subsection{T-matrix}

We will now turn to the calculation of the scaling function for the
T-matrix of the pseudogap Kondo problem,
along the crossover from the quantum critical into the local-moment regime.
Besides the comparison to NRG data, an important
question here is again the determination of the dynamical behavior, including the
low-frequency relaxational regime.
We will see that some control of the perturbation theory can be achieved due to the
peculiar logarithmic nature of the finite-temperature corrections in the
frequency-dependent T-matrix.
%In particular, the second expansion that we will consider in section~\ref{anderson}
%(done around the strong coupling fixed point) does not display a similar
%structure, and actually fails to grasp the low frequency limit of the T-matrix.
%NOT CLEAR!!

The T-matrix is identically zero at the decoupled LM fixed point, and starts at
second order in the Kondo coupling, with the result:
\begin{eqnarray}
\T^{(0)}(i\omega_n)=i \; \mr{sgn}(\omega_n) |\omega_n|^r J^2 N_0
\int_{-\infty}^{\infty} \!\!\! \textrm{d}x \frac{|x|^r}{1+x^2}
\end{eqnarray}
Note that the bare coupling $J$ appears here. This implies that this prefactor
will {\it not} follow an RG flow within the renormalized CS formulation we are
using, and simply acts as an non-universal prefactor.
For simplicity, we will omit the prefactor $J^2 N_0$ in the intermediate
steps of the following calculation, it will be re-instated in the final result.
We now want to calculate the full crossover function for this
quantity, and start by setting up the Callan-Symanzik formalism.

\subsubsection{Callan-Symanzik equations for $\mcal{T}$}
\label{app1}

The basic starting point is as previously\cite{bgz}
\begin{eqnarray}
\T^B(\omega,T,J,D)=Z_{\T}\T(\omega,T,j,\mu)
\end{eqnarray}
where $\T^B$ denotes the bare T-matrix, and $\T$ the renormalized T-matrix.
The renormalization factor is\cite{kircan}
\begin{equation}
Z_\T = 1-2j/r+O(j^2)
\end{equation}
at one-loop order. This gives
\begin{equation}
\eta_{\T}(j)=\beta(j)\frac{\partial \ln{Z}_{\T}}{\partial j} = -2j+O(j^2)\,.
\end{equation}
A general derivation of $Z_\T$ is also given in App.~\ref{apptmatrix}, which
shows the relation $Z_\T = Z_j^{-2} Z_f^2$, so that the bare coupling can also be
expressed as $J=\mu^{-r} Z_\T^{-1/2}N_0^{-1}j$. Since $\mu \partial J/\partial\mu=0$,
one gets the relation $\eta_\T(j) = -2r+\beta(j)/j$. Although $\eta_\T(j)$ picks
up contributions to all orders in perturbation theory, the existence of a
critical point $j_c$ such that $\beta(j_c)=0$ implies\cite{kircan} the {\it exact} result
$\eta_\T\equiv\eta_\T(j_c) = -2r$. This leads to $\T(\w)\sim|\w|^{-r}$ behavior
at the quantum critical point, as we will see below.

\begin{figure}[t!]
\begin{center}
\includegraphics[width=8cm]{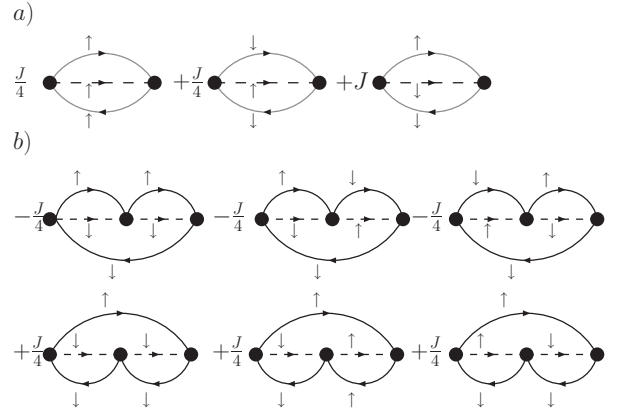}
\end{center}
\caption{
Diagrammatic expansion of the T-matrix of the Kondo model
up to next-to-leading order ($J^3$), with conventions given in Fig.~\ref{chiim}.
}
\label{Tmatrix}
\vspace{0.5cm}
\end{figure}

The diagrammatic expression for the T-matrix at order $j$ is shown in Fig.~\ref{Tmatrix}
including all spin indices. On a technical level it is crucial to perform the
perturbative expansion for the propagator of the complete operator
$T_{\sigma}=\sum_{\sigma'} \vec{S} \cdot
\vec{\tau}_{\sigma \sigma'} c^{\phantom{\dagger}}_{\sigma'}$, e.g.
$T_\uparrow = (1/2)(f^{\dagger}_\uparrow f^{\phantom{\dagger}}_\uparrow -
f^{\dagger}_\downarrow f^{\phantom{\dagger}}_\downarrow)\,
c^{\phantom{\dagger}}_\uparrow
+ f^{\dagger}_\downarrow f^{\phantom{\dagger}}_\uparrow
c^{\phantom{\dagger}}_\downarrow$.
If only the last term of the expression is chosen, namely $f^{\dagger}_\downarrow
f^{\phantom{\dagger}}_\uparrow c^{\phantom{\dagger}}_\downarrow$, one is
actually computing a correlation function which is not associated to a physical
observable, and one encounters double counting problems, which lead to an incorrect
result for $Z_\T$.

The same steps already followed to obtain the CS equations for the local
susceptibility give now:
\begin{eqnarray}
\T(\omega,T,j)&=&f_\T(\lambda) \lambda^r
\T\left(\frac{\omega}{\lambda},\frac{T}{\lambda},j(\lambda)\right)
\label{CS_T}
\end{eqnarray}
using that the bare scaling dimension of $\T$ is $r$ (hence the $\lambda^r$
prefactor). The anomalous behavior is provided by the function:
\begin{eqnarray}
f_\T(\lambda)&=&
\exp\left(\int_1^{\lambda}\frac{d\lambda'}{\lambda'}\eta_{\T}\left(j(\lambda')\right)
\right)\\
%& = & \left(\frac{r-j}{r}\right)^2 \left[\frac{1}{1+|\lambda\mu/T^\star|^r}\right]^2
& = & \left[\frac{\mu^r}{(T^\star)^r+|\lambda\mu|^r}\right]^2
\label{f_T}
\end{eqnarray}
which reduces to $f_\T(\lambda) = \lambda^{-2r}$ for $j=r$.
We now follow the RG philosophy by calculating the r.h.s. of equation~(\ref{CS_T})
in renormalized perturbation theory.

The diagrammatic expansion seen in Fig.~\ref{Tmatrix} gives the result for the
imaginary part of the T-matrix at order $j$:
\begin{eqnarray}
-\T''(\w,T,j) = |\w|^r\left [ 1 - 2 j \int dx |x|^r P\frac{x}{x^2-1}
\right. \nonumber \\
\left. \left(
\frac{|\w/\mu|^r}{\exp(-x|\w|/T)+1}-\theta(x)\right)\right ],
\end{eqnarray}
where $P$ denotes the principal value. In this formula, the pole coming from the
$Z_\T$ factor was taken into account, and appears as the $\theta$ term inside
the integral. For $\w=\mu$, it regularizes the expression in the UV so
that $\T_{pert}(\w=\mu)$ is finite. However, to make sense of this
formula when $\w\neq \mu$, one has to do a small $r$ expansion (up to order $r^0$),
leading to a cancellation of all $1/r$ poles, and to the result:
\begin{eqnarray}
\label{Tfinal}
\hspace{-1.5cm} -\T''(\w,T,j) & = & |\w|^r
\left[ 1 + 2j \ln\frac{|\w|}{\mu} - 2j I(\w/T) \right]\\
\nonumber
\mr{with} \;\; I(\w/T) & = & \int dx P\frac{x}{x^2-1} \\
\label{Iint}
&& \hspace{0.2cm} \left( \frac{1}{\exp(-x|\w|/T)+1}-\theta(x)\right)
\end{eqnarray}
(for consistency, we have also expanded the $|x|^r$ term to lowest order).

Using the CS equations, we will investigate in turn the zero-temperature crossover
regime ($T=0$, $j<r$), and then the finite-temperature critical regime ($T>0$, $j=r$),
the latter both at large and small frequency, corresponding to
critical and relaxational dynamics, respectively.

\subsubsection{Zero-temperature scaling function for $\T$}

\begin{figure}[!t]
\begin{center}
\includegraphics[width=8.5cm]{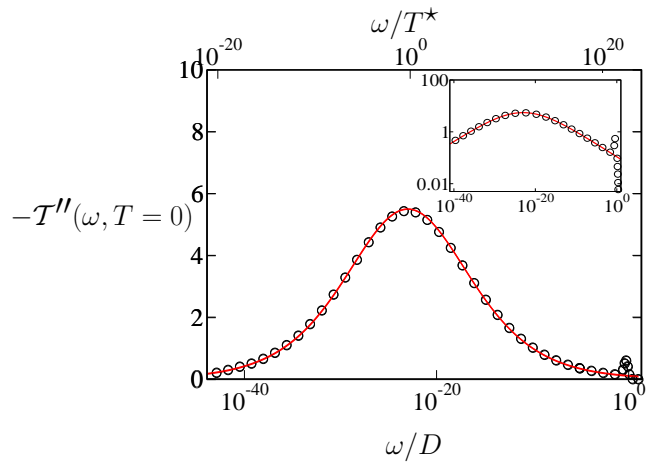}
\end{center}
\caption{(color online)
T-matrix $-\mcal{T}''(\w,T=0)$ at zero temperature for the same
parameters as in Fig.~\ref{TChi_r0.1}. NRG data point are the dots, and the
solid line follows formula~(\ref{Tmat_T0}), where the non universal amplitude
$J^2N_0\mu^{2r}$ has been fitted to match the numerics.
Note that the high-frequency peak at
$\w\sim U/2$ in the numerical result corresponds to charge fluctuations
(Hubbard bands) of the Anderson model employed in the NRG simulations, and cannot
be accounted for in our calculation based on the Kondo model.
The inset shows the same curve on a double-logarithmic scale.}
\label{comp}
\vspace{0.5cm}
\end{figure}

At zero temperature, the integral in Eq.~(\ref{Iint}) vanishes, and one is left
with a single frequency-dependent logarithmic divergency, similar to the
logarithmic temperature singularity in the perturbative expansion of the static
local susceptibility, formula~(\ref{line2}).
The application of the CS equation~(\ref{CS_T}) clearly imposes the choice
$\lambda=\w/\mu$ to eliminate this large term, which, using
Eqs.~(\ref{CS_T}-\ref{f_T}), gives:
\begin{equation}
-\T''(\omega,T=0,j<r) = J^2 N_0\, \frac{\w^r\mu^{2r}}{[(T^\star)^r+|\w|^r]^2}
\label{Tmat_T0}
\end{equation}
where we have re-introduced the non-universal prefactors.
This function crosses over from $\w^{r}/(T^\star)^{2r}$ behavior at $\w\ll T^\star$ to
$\w^{-r}$ behavior at $\w\gg T^\star$, as can be seen on
Fig.~\ref{comp}, and displays an excellent agreement with the NRG data.

\subsubsection{Finite-temperature scaling function for $\T$}

For simplicity, we first focus on the finite-temperature behavior of the T-matrix
above the quantum critical point, $j=r$.
At $T=0$, or when $\w\gg T$, the
previous arguments apply and the choice of the renormalization scale
$\lambda=\w/\mu$ together with the CS equation~(\ref{CS_T}) leads to the
critical behavior:
\begin{equation}
-\T''(\omega,T,j=r) \propto \frac{1}{\w^r} \;\; \mr{for} \;\; \w\gg T
\end{equation}
Note that the integral in~(\ref{Iint}) is of order 1 for $\w \geq T$
and provides only a small temperature correction to the leading critical
behavior, even when frequency is comparable to temperature.

We now investigate the low-frequency behavior of the finite-temperature
T-matrix. As discussed before (see Ref.~\onlinecite{sachdev} and Sec.~\ref{secchi}),
one generally obtains large terms in the expansion from perturbation theory,
preventing a controlled calculation.

Examining the integral~(\ref{Iint}) for $\w\ll T$, one clearly sees that
a $\ln(\w/T)$ divergency is pulled out, and combines with the already present
logarithm to give:
\begin{equation}
-\T''(\w,T,j) = |\w|^r
\left[ 1+2j\ln(T/\mu) + \ldots \right]
\end{equation}
where $\ldots$ denote regular $\w/T$ corrections at small frequency.
This shows that indeed large terms are generated in the perturbative expansion
in the relaxational regime. However, the extra singularity is logarithmic, which
suggests that the RG could be used to resum it. Indeed, applying the CS
equation~(\ref{CS_T}), using the above perturbative result on the r.h.s., and
taking into account that $f_\T(\lambda) = \lambda^{-2r}$ at the quantum
critical point, we get:
\begin{equation}
-\T''(\w,T,j=r) = \frac{|\w|^r}{\lambda^{2r}}
\left[ 1+ 2 r \ln(T/\lambda\mu)+\ldots\right]
\end{equation}
Now clearly the good choice
for the scale $\lambda$ is
\begin{equation}
\lambda = T/\mu ,
\end{equation}
so that perturbation theory is controlled.
This in turn implies the following small $\w$ behavior:
\begin{equation}
-\T''(\w,T,j=r) \propto \frac{|\w|^r}{T^{2r}} \;\; \mr{for} \;\; \w\ll T
\label{Trelax}
\end{equation}
up to small perturbative corrections.
We have to remark here that, although the $\w^r$ frequency behavior for $\T(\w,T,j=r)$
at $\w\ll T$ is reminiscent of the zero-temperature result near the local moment fixed
point, i.e., Eq.~(\ref{Tmat_T0}) for $\T(\w,T=0,j<r)$ at $\w\ll T^\star$,
criticality is reflected in the anomalous temperature dependent prefactor in Eq.~(\ref{Trelax}).
However, the fact that both frequency dependencies are identical
appears to be the reason behind the success in computing the relaxational dynamics
for the frequency dependent T-matrix directly from the RG equations.

\begin{figure}[!t]
\begin{center}
\includegraphics[width=8cm]{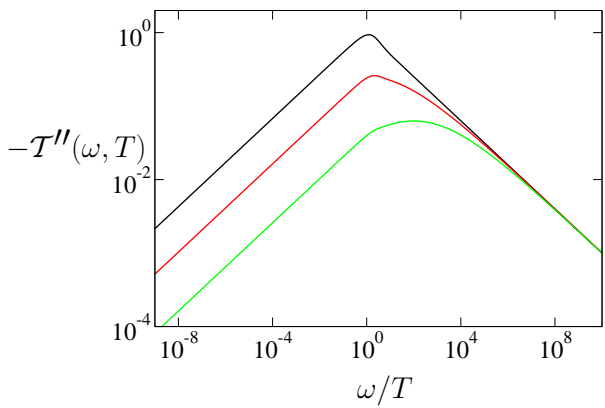}
\end{center}
\caption{(color online)
T-matrix $-\mcal{T}''(\w,T,j)$ for $r=0.3$ at finite temperature
from the formula~(\ref{Tfinal_T}) obtained via the analytic RG.
Different curves correspond to three values of $j<r$, associated from top to bottom
to $T^\star/T=0,1,100$ (the topmost curve corresponds thus to the quantum
critical point).
Note that no NRG data in the regime $\w<T$ is available to date.}
\label{Tmatrix_T}
\vspace{0.5cm}
\end{figure}

A generic expression for the critical crossover function can be further given by
noticing that the renormalized T-matrix, at the level of the CS equations, reads:
\begin{equation}
-\mcal{T}''(\w,T,j=r) = \frac{|\w|^r}{\lambda^{2r}}
\left[1+2j\ln\frac{|\w| e^{-I(\w/T)}}{\lambda\mu}\right]
\end{equation}
so that leading perturbative corrections vanish for the choice $\lambda=
e^{-I(\w/T)} |\w|/\mu$ [note that this alternative choice for $\lambda$
interpolates correctly between $T/\mu$ and $\w/\mu$ for both limits of small and
large $\w/T$]. At criticality, one thus has:
\begin{equation}
-\mcal{T}''(\w,T,j=r) = J^2 N_0 \frac{\mu^{2r}}{\w^{r}} e^{2rI(\w/T)}
\label{T_crit_T}
\end{equation}
which corresponds to the top curve shown in Fig.~\ref{Tmatrix_T}.
We now turn to the crossover regime ($j<r)$ at finite temperature. In this case,
the complete result at one-loop order reads:
\begin{equation}
-\T''(\omega,T,j<r) = J^2 N_0 \frac{\w^r\mu^{2r}}{[(T^\star)^r+|\w|^re^{-rI(\w/T)}]^2}
\label{Tfinal_T}
\end{equation}
which corresponds to the formula for the {\it full} scaling function
$\Phi_\T(\w/T,T/T^\star)$ defined in~(\ref{hyper_T}).

%\begin{eqnarray}
%\nonumber
%-\T''(\w,T,j) & = & |\w|^r \left[\frac{1}{1+|\Omega/T^\star|^r}\right]^2
%\left[ 1+ \frac{2r}{1+|\Omega/T^\star|^{-r}}
%\right. \\
%&& \left. \left( \theta(T-|\w|) \ln\frac{|\w|}{T} -I(\w/T) \right) \right]
%\end{eqnarray}
%where $\Omega=\mr{Max}(\w,T)$ and $I(\w/T)$ was defined in equation~(\ref{Iint}).
%
This function is plotted in Fig.~\ref{Tmatrix_T},
and we briefly comment on this result.
First, we notice the appearance of an asymmetric
hump for $\mcal{T}(\w,T,j)$ at $\w\sim T$, while such feature
is absent for $\mcal{T}(\w,T=0,j)$ at $\w\sim T^\star$ (see the inset on
Fig.~\ref{comp}). This is due to the presence of amplitude corrections
in the first case, formula~(\ref{Tfinal_T}), which are absent in the second
case, formula~(\ref{Tmat_T0}).
The shape of the finite-temperature scaling function is very reminiscent
of the result obtained from the local moment approach,\cite{logan} and a
quantitative comparison to these results could be interesting.
We note that our analytical estimate Eq.~(\ref{Trelax}) should apply
to the multichannel pseudogap Kondo model at large-$N$ as well.\cite{vojta}
As the T-matrix does not saturate for $\w<T$, but rather shows the
behavior~(\ref{Trelax}), we conclude that the conformal scaling ansatz
introduced for the metallic case~\cite{parcollet} does {\it not} hold
in general for the T-matrix.

However, from the considerations of Sec.~\ref{secchi} and Ref.~\onlinecite{sachdev},
it seems curious that a direct application of the RG equations for the real
frequency T-matrix does succeed. We want to show here that actually similar
(although less severe) problems do arise. While we believe that the frequency
behavior given by Eq.~(\ref{Tfinal_T}) is correct in the
$\w\rightarrow0$ limit, it turns out that sub-leading terms are incorrectly
given by the present calculation.
To see this clearly, we extend our results to the metallic case $r=0$.
When $\w\ll T$, the integral in~(\ref{Iint}) behaves as $I(\w/T) \simeq
\ln (\w/T) + \alpha_0 - \alpha_1 \w/T$ with $\alpha_0\simeq0.125$ and
$\alpha_1\simeq0.10$. Callan-Symanzik equations give in this case:
\begin{eqnarray}
-\T''(\omega,T,j>0) & = & \frac{1}{[\ln(\w/T_K)-I(\w/T)]^2} \\
& \simeq & \frac{1}{[\ln(T/T_Ke^{\alpha_0})+\alpha_1|\w/T|]^2}
\end{eqnarray}
This would imply both a shift in the Kondo temperature and a low-frequency
cusp, which are physically wrong. This type of artifact is related to the
difficulties which come from the use of RG equations for real frequency
quantities, as discussed previously. While they are quite severe in the metallic
Kondo problem, the fact that the finite temperature low-frequency T-matrix in the pseudogap case
generically vanishes as $\w^r$ at low frequency makes such corrections practically
hard to see. Indeed, at the critical point, formula~(\ref{T_crit_T}) implies an
incorrect prefactor $e^{2r\alpha_0}$, which leads to a negligibly small
overall amplitude correction to the exact result.

We thus conclude this section by
emphasizing that in the perturbative regime, $r\ll1$, all scaling functions of
the pseudogap Kondo problem have been computed with reasonable accuracy.

%%%%%%%%%%%%%%%%%%%%%%%%%%%%%%%%%%%%%%%%%%%%%%%%%%%%%%%%%%%%%%%%%%%%%%%

\section{Crossover functions near $r=1/2$:
Weak-coupling regime of the Anderson model}
\label{anderson}

We want here to investigate the vicinity of the quantum critical point SCR, now
on the side where the model flows to the strong coupling SSC fixed point, see
Fig.~\ref{flow}.
As detailed in Ref.~\onlinecite{fritz1}, we have to use the particle-hole symmetric Anderson
Hamiltonian as the effective model.
The expansion is performed around the non-interacting resonant-level model,
with the Coulomb interaction $U$ between the localized electrons being the
expansion parameter -- this allows a perturbatively controlled calculation
for DOS exponents $r$ close to 1/2.
The starting point is the action
\begin{eqnarray}
\mathcal{S}&=&-\frac{1}{\beta}\sum_{\omega_n,\s}\overline{d}_\sigma(\omega_n)\left[i
A_0 \mr{sgn}\left(\omega_n\right)|\omega_n|^r\right]d_{\sigma}(\omega_n)\nonumber \\
&+&\int_0^{\beta}\!\mr{d}\tau \; U\left(\overline{d}_\uparrow
d_\uparrow
-\frac{1}{2}\right)\left(\overline{d}_{\downarrow}d_\downarrow-\frac{1}{2}\right),
\label{phsymm}
\end{eqnarray}
where the local $d_\s$ fermions are dressed by conduction electrons ($\w_n$ 
above denote fermionic Matsubara frequencies $(2n+1)\pi/\beta$).
Technically, this expression can be obtained by integrating out the electronic
environment in~(\ref{HAnderson}), giving $A_0=\pi V^2 N_0
\cos^{-1}\left(\frac{\pi r}{2}\right)$.
The interaction term is written in an explicitly particle-hole
symmetric way.
We can define a dimensionless coupling $u$, related to the bare coupling $U$ by
$u=A_0^{-2} Z_4^{-1} Z \mu^{\epsilon} U$, where $\mu$ is the renormalization
energy scale, and we have absorbed the non-universal number $A_0$.
Further, $-\epsilon$ denotes the bare scaling dimension of the coupling
$U$, with $\epsilon=1-2r$.
The renormalization factors $Z_4$ and $Z$ are
associated to the vertex and field renormalization.
The beta function to two-loop order was explicitly derived in Ref.~\onlinecite{fritz1}:
\begin{eqnarray}
\beta(u)  =  \epsilon u-\AAA u^3 \,,~~
\AAA =  \frac{3(\pi-2 \textrm{ln}4)}{\pi^2}
\end{eqnarray}
with the fixed point $u_c^2 = \epsilon/\AAA$.
The renormalization factors read
\begin{equation}
Z_4 = 1 + \frac{\AAA u^2}{2\epsilon}\,,~Z = 1,
\end{equation}
the latter relation being exact.\cite{fritz1}

In a manner similar to Sec.~\ref{kondo} we now calculate the running coupling
constant, which is simply obtained by the substitution $\kappa=u^2$. We find
\begin{eqnarray}
u^2(\lambda)= \frac{\epsilon/\AAA}
{1+[(2\epsilon-u)/u]\lambda^{-2\epsilon}}
\end{eqnarray}
We can write this coupling in a similar manner as Eq.~(\ref{running_j}),
yielding for $u\leq u_c=\sqrt{\epsilon/\AAA}$:
\begin{eqnarray}
u^2(\lambda)=\frac{\epsilon/\AAA}{1+\left( \frac{\lambda\mu}{T^\star}\right)^{-2 \epsilon}}
\label{ut}
\end{eqnarray}
where the crossover scale from SCR to SSC is given by
\begin{equation}
T^\star=\mu \left(\frac{u^2}{u_c^{2}-u^2}\right)^{1/(2\epsilon)}.
\end{equation}
Note that the ``distance'' between SCR and SSC is of order $\sqrt{\epsilon}$
(and not $\epsilon$), which makes the present expansion numerically less accurate
than the one performed from the Kondo model.

\subsection{Impurity susceptibility}

\begin{figure}[!t]
\begin{center}
\includegraphics[width=7.8cm]{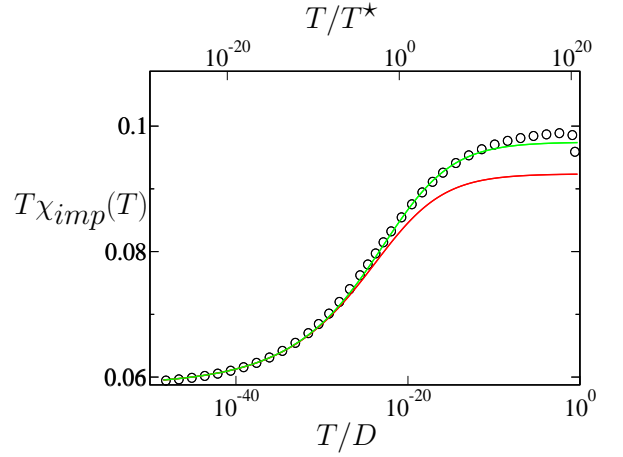}
\end{center}
\caption{(color online)
$T\Cimp(T)$ at $r=0.47$ capturing the crossover
between SCR at high temperature and SSC at low temperature.
The NRG data (dots) are obtained for $U/D=1$, $V/D=0.845$, while the analytical
curves are plotted according to the perturbative result, respectively the full
one-loop result~(\ref{TChi1}) for the upper curve, and the strict
$O(\sqrt\epsilon)$ result~(\ref{TChi2}) for the lower curve (see text). The
extracted crossover scale is $T^\star/D\simeq1\times10^{-21}$.
Note that the agreement is as always better at low temperature where the calculation 
is perturbatively controlled.}
\label{TChi_r0.47}
\vspace{0.5cm}
\end{figure}

We start by describing the crossover in the static quantities.
It is worthwhile noting that although the
critical point SCR is identical to the one found using the weak-coupling expansion
within the Kondo model, the crossover is now different, since we here expand around the
strong-coupling fixed point of the Kondo model.
Focusing on the impurity susceptibility, and using the perturbative expression
derived in Ref.~\onlinecite{fritz1} together with the running coupling~(\ref{ut}),
we find that
\begin{eqnarray}
\label{TChi1}
T \Cimp(T) & = & \frac{r}{8} + \frac{(1-2^{-1-r})^2 \zeta(1+r)^2}{2 \pi^{2+2r}}
u(T/\mu)\\
& \simeq & \frac{r}{8} + 0.19 \sqrt{\frac{1}{2}-r}
\frac{1}{\sqrt{1+\left(\frac{T}{T^\star} \right)^{-2\epsilon}}}.
\label{TChi2}
\end{eqnarray}
where $r\simeq1/2$ has been inserted in the prefactor of eq.~(\ref{TChi1}) to obtain the 
last equation. This result is compared to the numerical data obtained by NRG on
Fig.~\ref{TChi_r0.47}. We have to note that the present expansion also differs on a
quantitative level from the one based on the Kondo model in the sense that the
$r$-dependent prefactors, such as the one in equation~(\ref{TChi1}), vary quite
rapidly with $r$ (this is not the case for the coefficient in
equation~(\ref{chiloccor})). Although a {\it strict} expansion at order $\sqrt\epsilon$
requires to expand the prefactor at order $\epsilon^0$ as in
equation~(\ref{TChi2}), which gives the lower curve of figure~\ref{TChi_r0.47},
a much better agreement in the range $r=0.4\ldots0.5$ can be reached by using
the full value of the $r$-dependent prefactor, as seen in the upper curve of 
figure~\ref{TChi_r0.47}.

As in Sec.~\ref{entropy}, the impurity entropy has tiny variations, and we do not
study this quantity further.

\subsection{Local susceptibility}

Similarly to the calculation done with the Kondo model, we write down the CS equation for
the local susceptibility in the Anderson model. The renormalization factor $Z_{\chi}$
was calculated in Ref.~\onlinecite{fritz1}, with the result
\begin{equation}
Z_{\chi}=1-\frac{2u}{\pi \epsilon}
\end{equation}
giving the anomalous exponent
\begin{equation}
\eta_\chi'(u)=-\frac{2}{\pi}u
\end{equation}
Note that $\eta_\chi'$ differs from $\eta_\chi$ defined previously as the
expansion is done now around a resonant level and not around a free spin limit.
These coefficients are related by the relation $1-\eta_\chi=-\epsilon-\eta_\chi'$, 
correcting a misprint in Ref.~\onlinecite{fritz1}.
The scaling equation for the local susceptibility within the Anderson model
naturally reads:
\begin{eqnarray}
\Cloc(i \nu,T,u) & = & \lambda^{\epsilon} g_\chi(\lambda)
\Cloc \left(\frac{i\nu}{\lambda},\frac{T}{\lambda},u(\lambda)\right)\\
\mr{with} \;\; g_\chi(\lambda) & = & \exp\left(\int_u^{u(\lambda)}
\!\!\textrm{d}u'\; \frac{\eta_\chi'(u')}{\beta(u')}\right)
\end{eqnarray}
using the fact that $\Cloc(i \nu,T,u)$ has bare scaling dimension
$\epsilon$. The exponential prefactor can be evaluated to:
\begin{eqnarray}
g_\chi(\lambda) = \left[\frac{\left(u(\lambda)+\sqrt{\epsilon/\AAA} \right)
\left(u-\sqrt{\epsilon/\AAA}\right)}{\left(u(\lambda)-\sqrt{\epsilon/\AAA} \right)
\left(u+\sqrt{\epsilon/\AAA}\right)}\right]^{\frac{1}{\pi \sqrt{\epsilon/\AAA}}}
\end{eqnarray}

\begin{figure}[!t]
\begin{center}
\includegraphics[width=7.0cm]{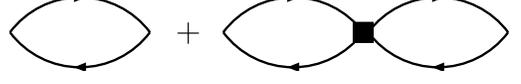}
\end{center}
\caption{
Diagrammatic contribution to the local susceptibility of the Anderson
model up to order $u$, where full lines denote the propagator of the hybridized
$d$ level, and squares the $U=\mu^{-\epsilon}A_0^2u$ vertex.}
\label{chiloc1}
\end{figure}

The perturbative expansion for the local susceptibility is given in
diagrammatic form in Fig.~\ref{chiloc1}, and simply reads at lowest
order in $u$:
\begin{equation}
\Cloc(i\nu_n) = Z_\chi^{-1} [\chi_0(i\nu_n) + u \mu^{-\epsilon} \chi_0(i\nu_n)^2]
\label{chiand}
\end{equation}
In the above expression, $\chi_0$ is the susceptibility of the resonant level,
and involves the computation of a simple bubble graph.
(Again, the renormalization factor $Z_4$ is unity to the order needed here.)
Focusing first on the $T=0$ case, we find
$\chi_0(i \nu) = -|\nu|^{\epsilon}/(\pi\epsilon)$ as $\epsilon\ll1$.
We remark that a direct evaluation of the $Z_\chi$ factor from Eq.~(\ref{chiand})
and the above expression would provide the {\it incorrect} result
$Z_\chi = 1-u/(\pi\epsilon)$.
This is because $\chi_0(i \nu)$ presents already a pole
at bare level, invalidating the estimation of $Z_\chi$. To do things in the
proper way, it is better to perform the analytic continuation in
Eq.~(\ref{chiand}) and take its imaginary part:
\begin{equation}
\Cloc''(\w) = Z_\chi^{-1}
\chi_0''(\w)\left[1+2u\mu^{-\epsilon}\chi_0'(\w)\right]
\label{extra}
\end{equation}
where the real and imaginary parts of the bare susceptibility at $T=0$ are given by the
expressions
\begin{eqnarray}
\chi_0'(\omega,T=0)&=&-\frac{|\omega|^\epsilon}{\pi \epsilon}
\cos(\pi\epsilon/2) \simeq -\frac{|\omega|^\epsilon}{\pi \epsilon} \\
\nonumber
\chi_0''(\omega,T=0)&=&\frac{\sin(\pi\epsilon/2)}{\pi\epsilon/2}
\frac{|\omega|^\epsilon}{2} \textrm{sgn}\left( \omega \right) \simeq
\frac{|\omega|^\epsilon}{2} \textrm{sgn}\left( \omega \right)\\
\label{chi0i}
\end{eqnarray}
(above $\epsilon\ll1$ has been used to simplify expressions).

Note that now $\chi_0''(\omega,T=0)$ is well defined in the small $\epsilon$
limit and that an extra factor $2$ appears in~(\ref{extra}) with respect
to~(\ref{chiand}), which is crucial to finally obtain the
correct $Z_\chi=1-2u/(\pi\epsilon)$ factor.

\begin{figure}[!b]
\begin{center}
\includegraphics[width=8cm]{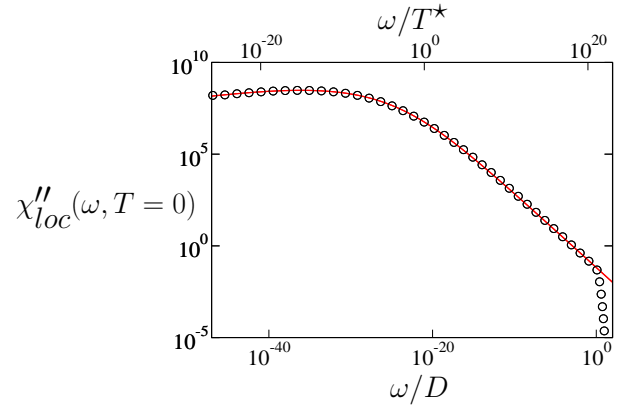}
\end{center}
\caption{(color online)
Imaginary part of the dynamic local susceptibility
$\chi_{loc}''(\w,T=0)$ at zero temperature for $r=0.47$.
The analytical result is in Eq.~(\ref{cloc_r12}).}
\label{Chiloc_r0.47}
\vspace{0.5cm}
\end{figure}

Finally, the CS equation with the choice $\lambda = \w/\mu$ results in:
\begin{equation}
\Cloc''(\w,T=0,u) = \frac{1}{2} g_\chi(\w/\mu) |\w|^\epsilon \mr{sgn}(\w)
\label{cloc_r12}
\end{equation}
which crosses over from $\w^\epsilon$ behavior at $\w\ll T^\star$ to
$\w^{\epsilon -(2/\pi) \sqrt{\epsilon/\AAA}}$ behavior at $\w\gg T^\star$,
in agreement with the anomalous exponent value\cite{errors}
\begin{equation}
\eta_\chi' = - \frac 2 \pi \frac \epsilon \AAA .
\end{equation}
The above complete expression (\ref{cloc_r12}) is compared to NRG results in
Fig.~\ref{Chiloc_r0.47}.

Next we turn to the finite-temperature behavior of the susceptibility.
As opposed to the case of the Kondo model where $\Cloc$ starts only at order $j^2$,
the non-interacting $u=0$ limit already shows a non-trivial result for
$\Cloc$, that includes a temperature dependence as well.
Computing the bare bubble at $T>0$, we indeed find:
\begin{eqnarray}
\chi_0'(\omega,T)&=&-\frac{T^\epsilon}{2\pi}
\int_0^{+\infty} \!\!\!\!\!\! dx \;
x^{-r} \tanh\left(\frac{x}{2}\right)\\
\nonumber
&& \left[\,\left|\frac{\w}{T}-x\right|^{-r}\mr{sgn}\left(\frac{\w}{T}-x\right)
-\left|\frac{\w}{T}+x\right|^{-r}\right]\\
\chi_0''(\omega,T)&=&\frac{T^\epsilon}{2\pi}
\int_0^{+\infty} \!\!\!\!\!\! dx \;
x^{-r} \tanh\left(\frac{x}{2}\right) \\
\nonumber
&& \left[\,\left|\frac{\w}{T}-x\right|^{-r}
-\left|\frac{\w}{T}+x\right|^{-r}\right]
\end{eqnarray}
This expression recovers the zero-temperature result~(\ref{chi0i}) when $\w\gg T$,
and gives in the opposite limit:
\begin{eqnarray}
\chi_0'(\omega\ll T) &=& -\frac{T^\epsilon}{\pi\epsilon} +
T^\epsilon \log\frac{2e^\gamma}{\pi} \\
\chi_0''(\omega\ll T) &=& \frac{T^\epsilon}{2\pi\epsilon} \frac{\w}{T}
\end{eqnarray}
Applying the CS equation in this regime (for simplicity we focus on the critical
point $u=u_c$) gives for the susceptibility (choosing $\lambda=T/\mu$):
\begin{equation}
\chi_{loc}''(\w\ll T) = T^\epsilon \left(\frac{T}{\mu}\right)^{-(2/\pi)\sqrt{\epsilon/\AAA}}
\!\!\! \frac{1}{2\pi\epsilon} \frac{\w}{T} [1+2u_c \log\frac{2e^\gamma}{\pi}]
\end{equation}
which is the expected relaxational behavior. In the present case, the success of
the RG is due to the fact that the {\it bare} susceptibility already presents the
correct small $\w/T$ limit.

\subsection{T-matrix}

In the Anderson model, the T-matrix is simply proportional to the propagator
$G_d$ of the local $d$ level.
In the non-interacting limit ($u=0$) one has
the free propagator $G_{d0}(i\w_n) = [i\mr{sgn}(\w_n)|\w_n|^r]^{-1}$
whose analytic continuation reads:
\begin{eqnarray}
\nonumber
G_{d0}'(\omega,T)&=&\sin(\pi r/2) |\w|^{-r} \mr{sgn}(\w) \simeq
\frac{1}{\sqrt{2}} |\w|^{-r} \mr{sgn}(\w)\\
G_{d0}''(\omega,T)&=& - \cos(\pi r/2) |\w|^{-r} \simeq
-\frac{1}{\sqrt{2}} |\w|^{-r}
\end{eqnarray}
using $r\simeq1/2$ to simplify expressions.
Clearly, the T-matrix obeys $\T(\omega) \propto \frac{1}{\omega^r}$ at tree level
(i.e. at the SSC fixed point). Since we have already seen in the study of the
pseudogap Kondo model that the same behavior obtains at the quantum critical
point, we should expect the absence of an anomalous exponent
(see Ref.~\onlinecite{fritz1} for a proof of this statement), and simple amplitude
corrections for the crossover between SCR and SSC
(see also Ref.~\onlinecite{bulla}).

\begin{figure}[!t]
\begin{center}
\includegraphics[width=2.0cm]{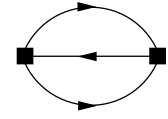}
\end{center}
\caption{Self-energy of the Anderson model at order $u^2$.}
\label{self}
\end{figure}

\begin{figure}[!t]
\begin{center}
\includegraphics[width=8.3cm]{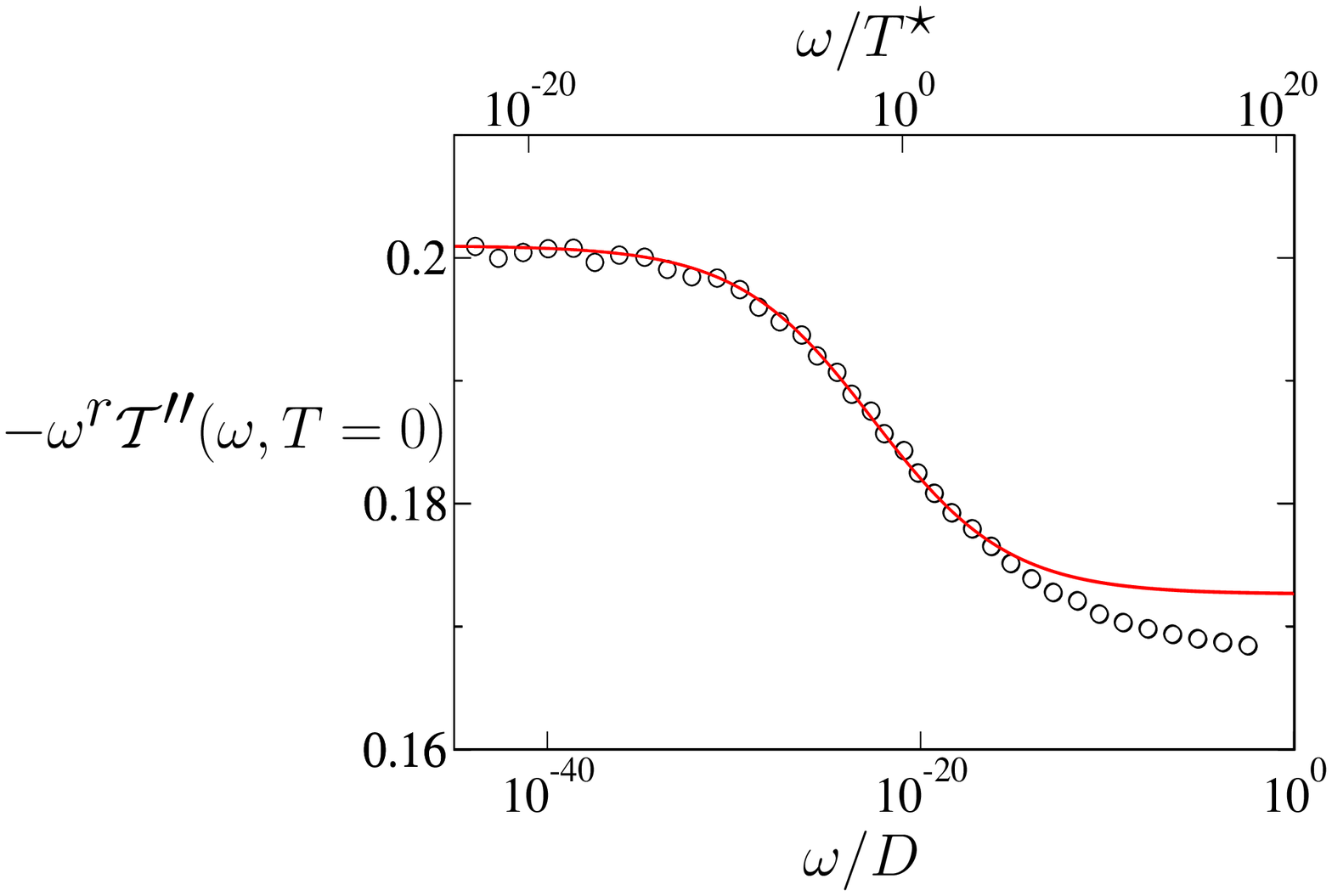}
\end{center}
\caption{(color online)
Rescaled T-matrix $-\w^r\mcal{T}''(\w,T=0)$, Eq.~(\ref{tmat_r12}), at zero
temperature for the same parameters as in Fig.~\ref{TChi_r0.47}, showing the
crossover on the amplitude of the common power law $\w^{-r}$.
The ringing at low frequency in the NRG data
comes from numerical errors in the calculated ratio related to the
broadening procedure.
}
\label{extract_rhof_r0.47}
\vspace{0.5cm}
\end{figure}

The corrections to the bare propagator brought about by the local Coulomb
interaction $u$ are simply expressed by the self-energy, displayed in
Fig.~\ref{self}, which at zero temperature amounts to (for $r\simeq1/2$):
\begin{eqnarray}
\Sigma_d(\tau) & = & U^2 G_{d0}(\tau)^3\\
& = & \mu^{-2\epsilon} u^2
\left[-\frac{1}{\pi}
\frac{\cos(\pi r/2) \Gamma(1-r)}{|\tau|^{1-r}\mr{sgn(\tau)}}\right]^3 \\
& \simeq & -\mu^{-2\epsilon} \frac{1}{(2\pi)^{3/2}} \frac{1}{|\tau|^{3-3r}}
\mr{sgn}(\tau)
\end{eqnarray}
The Fourier-transformed self-energy reads $\Sigma(i\w) =
-i\mu^{-2\epsilon} (u^2/\pi) |\w|^{2-3r} \mr{sgn}(\w)$, so that the propagator is:
\begin{eqnarray}
G_d^{-1}(i\w) & = & G_{d0}^{-1}(i\w)-\Sigma_d(i\w) \\
& = & i |\w|^r \mr{sgn}(\w) \left[1+\frac{u^2}{\pi}
\left|\frac{\w}{\mu}\right|^{2\epsilon}\right]
\end{eqnarray}
The analytic continuation gives for the perturbative propagator at
order $u^2$
\begin{equation}
G_d''(\w,T=0) = -2\sqrt{2} |\w|^{-r} \frac{1}{1+u^2/\pi} \,.
\end{equation}
We are now in the position to apply the CS equation with $\lambda=\w/\mu$
(there is no multiplicative term here as $Z_\T=1$), resulting in:
\begin{equation}
\mcal{T}''(\w,T=0) \propto |\w|^{-r} \left[1+\frac{\epsilon}{\pi \AAA}
\frac{1}{1+|\w/T^\star|^{-2\epsilon}}\right]^{-1} \,.
\label{tmat_r12}
\end{equation}
This expression is plotted in Fig.~\ref{extract_rhof_r0.47}, together
with re-scaled NRG data showing reasonably good agreement.

We finally conclude on the finite-temperature behavior of the T-matrix. From the
study of the Kondo model, we know that it should obey $\mcal{T}''(\w,T,u_c) \sim
|\w|^r/T^{2r}$ for $\w\ll T$ at the quantum critical point.
However, we start at tree level with a
$|\w|^{-r}$ dependence with $r\simeq 1/2$, so that the use of CS equations on the
real-frequency axis is bound to fail, as one would need a large (hence non-perturbative)
correction of the exponent.
At this stage it is unclear whether the
strategy of using the CS equation directly in imaginary time may be useful here.

%%%%%%%%%%%%%%%%%%%%%%%%%%%%%%%%%%%%%%%%%%%%%%%%%%%%%%%%%%%%%%%%%%%%%%%

\section{Conclusion}

In this paper we have undertaken a detailed study of the crossovers that occur
in the particle-hole symmetric pseudogap Kondo and Anderson models, by computing
various universal scaling functions associated to the magnetic and electronic
response.
Our analytical calculations are based on the use of the perturbative
renormalization group and Callan-Symanzik equations, a powerful method that was
not previously pursued in its generality for quantum phase transitions.
The comparison of our results to the available numerical data obtained by NRG
simulations was seen to be impressive, validating the present approach.
Some predictions for the low-frequency limit of dynamic quantities at finite
temperature were also made, a regime which remains challenging for the numerics.

\begin{figure}[!t]
\begin{center}
\includegraphics[width=7.8cm]{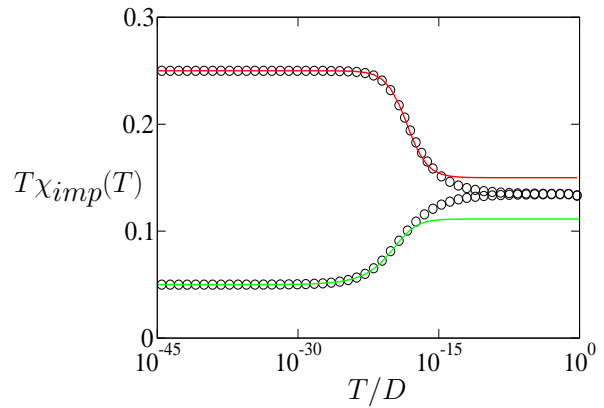}
\end{center}
\caption{Impurity susceptibility $T\Cimp(T)$ for $r=0.4$. The upper curve is obtained from the 
Kondo expansion result Eq.~(\ref{tchiimp}), while the lower curve originates
from the Anderson one, Eq.~(\ref{TChi2}). The NRG data (dots) are taken on both
sides of the quantum critical point, illustrating flow towards the local moment
and the resonant level fixed points respectively.}
\label{global}
\end{figure}

As a final illustration, we have compared in Fig.~\ref{global} the NRG data to our 
lowest-order estimates for the impurity susceptibility, {\it both} for small $r$ and 
small $1/2-r$, taking a common value of $r=0.4$ which lies beyond the limit
where our expansions are quantitatively accurate. This nevertheless illustrates in a 
very nice manner how two seemingly different expansions are able to capture both
sides of a common quantum critical point.

Finally, we think that the formalism that we developed and illustrated on these
peculiar quantum impurity models will be useful for a wider class of problems
showing quantum criticality, e.g., for impurity models with bosonic baths and
for bulk quantum magnets.

%%%%%%%%%%%%%%%%%%%%%%%%%%%%%%%%%%%%%%%%%%%%%%%%%%%%%%%%%%%%%%%%%%%%%%%

\acknowledgments

We thank M. Kir\'{c}an and H. Kroha for helpful discussions, G. Zar\'and for
useful advice, particularly regarding the derivation of the T-matrix of the
Kondo model, and finally R. Bulla and T. Pruschke for help with the NRG calculations.
This research was supported by the Deutsche Forschungsgemeinschaft through
the Center for Functional Nano\-structures (CFN) and the
Virtual Quantum Phase Transitions Institute in Karlsruhe.

%%%%%%%%%%%%%%%%%%%%%%%%%%%%%%%%%%%%%%%%%%%%%%%%%%%%%%%%%%%%%%%%%%%%%%%

\appendix

\section{Field-Theoretic derivation of the T-matrix}
\label{apptmatrix}

In this Appendix we derive the expression for the T-matrix of the Kondo model
within a path integral representation and provide a proof of the formula
$Z_\T=Z_f^2Z_j^{-2}$ for the associated renormalization factor.
Following standard literature in field theory~\cite{zinn},
we can express the single-particle matrix elements of the T-matrix using
reduction formulas to the full Green function. Because of the local nature of
the interaction, this can be cast in the simple form~\cite{zarand2,zarand3}
\begin{equation}
G_c(i\omega_n,k,k')= G_{c0}(i\omega_n,k)[1+\T(i \omega_n) G_{c0}(i\omega_n,k')]
\label{deftmatrix}
\end{equation}
where $G_{c0}$ stands for the bath Green function of the free system, whereas $G_c$
represents the interacting system. We will start by writing out the
generating functional for the fully interacting Green function of the
electrons in the Kondo problem~\cite{negele}
\begin{eqnarray}
Z(\eta_\sigma, \overline {\eta}_\sigma)= \int D \left[\overline{c}_{ \sigma} c_{\sigma} \right]
D\left [ \overline{f}_{\sigma} f_{\sigma} \right]e^{-S},
\end{eqnarray}
where the full action of the problem reads:
\begin{eqnarray}
\mathcal{S}&=&-\frac{1}{\beta} \sum_{i \omega_n} \int \textrm{d}^dk \;
\overline{c}_{\sigma}(i\omega_n, k) G^{-1}_{c0}(i\omega_n,k)c_{\sigma}(i\omega_n,k)\nonumber \\
&-& \frac{1}{\beta}\sum_{i \omega_n}
i\w_n \overline{f}_{\sigma}(i\w_n) f_{\sigma}(i\omega_n)\nonumber \\
&+&J \int \textrm{d}\tau \overline{f}_{\sigma}
\frac{\vec{\tau}_{\sigma \sigma'}}{2}f_{\sigma'} \cdot \overline{c}_{\alpha}(0)
\frac{\vec{\tau}_{\alpha \beta}}{2} c_{\beta}(0) \nonumber \\
&-& \frac{1}{\beta}\sum_{i\omega_n} \int \textrm{d}^d k
\left[ \overline{\eta}_\sigma c_{\sigma}(i\omega_n,k)+ \textrm{h.c.}\right]
\end{eqnarray}
Using this generating functional we obtain the fully interacting Green function by derivation
\begin{eqnarray}
\frac{\delta^2 \textrm{ln} Z(\overline{\eta},\eta)}{\delta \overline{\eta}_{\sigma}(i \omega_n,k) \delta
\eta_\sigma(i\omega_n,k')} \, \rule[-8pt]{0.3pt}{18pt}_{\,
\eta_\sigma=\overline{\eta}_\sigma=0}= G_c(i\omega_n,k,k').\nonumber \\
\end{eqnarray}
performing the shift
$c_{\sigma}\rightarrow c_{\sigma}+[G_{c0}\eta_\sigma]$ (this is understood as a
product in $\w_n$ space or a convolution in $\tau$ space), we arrive at:
\begin{eqnarray}
\mathcal{S}&=&-\frac{1}{\beta} \sum_{i \omega_n} \int \textrm{d}^dk
\overline{c}_{\sigma}(i\omega_n, k)
G^{-1}_{c0}(i\omega_n,k)c_{\sigma} (i\omega_n,k)\nonumber \\
&-& \frac{1}{\beta} \sum_{i \omega} \int \textrm{d}^d k
\overline{\eta}_\sigma(i \omega_n,k)G_{c0}(i\omega_n, k)\eta_\sigma(i\omega_n,k) \nonumber\\
&-& \frac{1}{\beta}\sum_{i \omega_n} i\w_n
\overline{f}_{\sigma}(i\omega_n)f_{\sigma}(i\omega_n)\nonumber \\
&+&J \int \textrm{d}\tau \overline{f}_{\sigma}
\frac{\vec{\tau}_{\sigma \sigma'}}{2}f_{\sigma'}
\cdot \overline{c}_{\alpha}(0) \frac{\vec{\tau}_{\alpha \beta}}{2}
c_{ \beta}(0) \nonumber \\
&+&J \int \textrm{d}\tau \overline{f}_{\sigma}
\frac{\vec{\tau}_{\sigma \sigma'}}{2}f_{\sigma'} \cdot [\overline{\eta}_{\alpha}(0)G_{c0}]
\frac{\vec{\tau}_{\alpha \beta}}{2} c_{ \beta}(0) \nonumber \\
&+&J \int \textrm{d}\tau \overline{f}_{\sigma}
\frac{\vec{\tau}_{\sigma \sigma'}}{2}f_{\sigma'} \cdot \overline{c}_{\alpha}(0)
\frac{\vec{\tau}_{\alpha \beta}}{2} [\eta_{ \beta}(0)G_{c0}] \nonumber \\
&+&J \int \textrm{d}\tau \overline{f}_{\sigma}
\frac{\vec{\tau}_{\sigma \sigma'}}{2}f_{\sigma'} \cdot [\overline{\eta}_{\alpha}(0)G_{c0}]
\frac{\vec{\tau}_{\alpha \beta}}{2} [\eta_{ \beta}(0)G_{c0}]\nonumber\\
\end{eqnarray}
\begin{figure}
\begin{center}
\includegraphics[width=6cm]{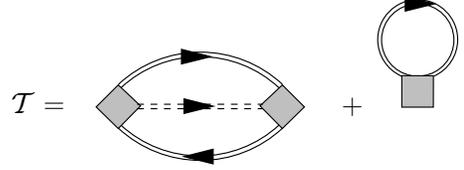}
\end{center}
\caption{
T-matrix in diagrammatic form:
The double solid line is the full impurity local propagator of the Abrikosov fermions
$f_\sigma$, the double dashed line represents the full local propagator of the conduction
electrons $c_\sigma$, and the filled square denotes the full vertex function.
The second graph vanishes in absence of a magnetic field.
}
\label{tmatrixdiagrams}
\vspace{0.5cm}
\end{figure}
Taking the derivative of this expression w.r.t. the source terms $\eta$ and
$\overline{\eta}$, we find
the following expression for the fully interacting single-particle Green
function:
\begin{eqnarray}
G_c(i\omega,k,k')&=&G_{c0}+ J^2 G_{c0} \langle T^{\phantom{\dagger}}_\sigma(i\omega_n)
T^{\dagger}_\sigma(i\omega_n)\rangle G_{c0}\nonumber \\
&+&J G_{c0} \langle \sum_{\alpha \beta}f^{\dagger}_\alpha \frac{\vec{\tau}_{\alpha \beta}}{2}
f^{\phantom{\dagger}}_\beta \vec{\tau}_{\sigma \sigma} \rangle G_{c0}.
\end{eqnarray}
where we have introduced the composite fermionic operator $T_\sigma = \sum_{\sigma'
\alpha \beta} f^{\dagger}_\alpha \frac{\vec{\tau}_{\alpha \beta}}{2}
f^{\phantom{\dagger}}_\beta \vec{\tau}_{\sigma \sigma'} c_{\sigma'}(0)$.
Comparing this expression with (\ref{deftmatrix}), we can identify the T-matrix as
\begin{eqnarray}
\nonumber
\T(i\omega_n)&=&J^2\langle T^{\phantom{\dagger}}_\sigma(i\omega_n)
T^{\dagger}_\sigma(i\omega_n)\rangle
+J \langle \sum_{\alpha \beta}f^{\dagger}_\alpha \frac{\vec{\tau}_{\alpha \beta}}{2}
f^{\phantom{\dagger}}_\beta \vec{\tau}_{\sigma \sigma} \rangle\\
\label{T}
\end{eqnarray}
which is illustrated in Fig.~\ref{tmatrixdiagrams} in diagrammatic language.
The second graph, which corresponds to the second term in equation Eq.~\ref{T},
vanishes in the absence of a magnetic field.
It is now only a matter of identification to note that expression for $Z_\T$
is given by \cite{kircan} $Z_\T=Z_j^{-2} Z_f^2$.

%%%%%%%%%%%%%%%%%%%%%%%%%%%%%%%%%%%%%%%%%%%%%%%%%%%%%%%%%%%%%%%%%%%%%%%%%


\begin{thebibliography}{25}
\bibitem{nrgpara}
The NRG calculations have been performed with discretization
parameter $\Lambda = 2$, keeping $N_s = 800$ levels.
In the figures, only a fraction of the available NRG data points
are shown, to reduce the number to a few per decade in temperature
or frequency.
\bibitem{errors}
The Eq.~(47) for $\eta_\chi$ in Ref.~\onlinecite{fritz1}
contained a sign error, in the notation of Ref.~\onlinecite{fritz1}
it should read
$\eta_\chi = 2-2r-2.688 \sqrt{\frac{1}{2} -r}+ {\cal O}(\epsilon)$.
The corresponding Fig. 5 of Ref.~\onlinecite{fritz1} is correct.
\end{thebibliography}
\end{document}